\documentclass[aps,prb,
,floatfix,footinbib,longbibliography,
preprint
]{revtex4-1}
\usepackage{epsfig}
\usepackage{graphicx}
\usepackage{dcolumn}
\usepackage{bm}
\usepackage{mathrsfs}
\usepackage{amsmath}
\usepackage{bbold}
\usepackage{color}
\usepackage{epstopdf}
\usepackage{subfigure}
\usepackage{appendix}

\usepackage[colorlinks=true,pdfborder=001,linkcolor=blue,anchorcolor=blue,citecolor=blue,urlcolor=blue]{hyperref}

\begin{document}

\title{Nonequilibrium reservoir engineering of a biased coherent conductor for hybrid energy transport in nanojunctions}
\author{Bing-Zhong Hu}
\author{Lei-Lei Nian}
\email{llnian@hust.edu.cn}
\author{Jing-Tao L\"{u}}
\email{jtlu@hust.edu.cn}
\affiliation{School of Physics and Wuhan National High Magnetic Field Center, Huazhong University of Science and Technology, Wuhan 430074, P. R. China}
\date{23 March, 2019}

\begin{abstract}

We show that a current-carrying coherent electron conductor can be treated as effective bosonic energy reservoir involving different types of electron-hole pair excitation. For weak electron-boson coupling, hybrid energy transport between nonequilibrium electrons and bosons can be described by a Landauer-like formula. This allows for unified account of a variety of heat transport problems in hybrid electron-boson systems. As applications, we study the non-reciprocal heat transport between electrons and bosons, thermoelectric current from a cold-spot and electronic cooling of the bosons.
Our unified framework provides an intuitive way of understanding hybrid energy transport between electrons and bosons. It opens the way of nonequilibrium reservoir engineering for efficient energy control between different quasi-particles in the nanoscale.

\end{abstract}

\maketitle

\section{Introduction}

Understanding nonequilibrium energy transport in the nanoscale is of crucial importance both for the fundamental development of quantum thermodynamics and for the practical application of nanoscale thermal, thermoelectric and optoelectronic devices. For phase coherent transport, the celebrated Landauer-B\"uttiker formalism has been successfully applied to study quasi-particle energy transport following different statistics, including electrons\cite{imry1999conductance}, photons\cite{ojanen2008mesoscopic,biehs2010mesoscopic,zhang2018energy,benabdallah2014near}, phonons\cite{rego1998quantized,mingo2005carbon,yamamoto2006nonequilibrium,wang2006nonequilibrium,wang2007nonequilibrium,wang2008quantum,ruokola2009thermal,li2012colloquium,taylor2015quantum,wang2016landauer} and magnons\cite{wang2004spin}.
Wherein, the baths connecting to the system are assumed to be in thermal equilibrium with given temperature and/or chemical potential, where the quasi-particle distribution function  is determined by its statistics, i.e., the Fermi-Dirac distribution for fermions, and the Bose-Einstein distribution for bosons. A difference in the distribution drives an energy current flow between the two thermal baths.

However, the same approach is difficult to describe energy transport between quasi-particles following different statistics, which is ubiquitous in thermoelectric and optoelectronic processes of nano-junctions. Examples of such processes include electroluminescence\cite{kuhnke2017atomic,galeprin2017photonics,schneider2010optical,schneider2012light}, Joule heating\cite{huang2007local,ioffe2008detection,lu_current-induced_2015,hartle2011resonant,hartle2011vibrational,hartle2018cooling}, current-induced\cite{galperin2009cooling,simine2012vibrational,lykkebo2016single,hartle2011resonant} or radiative cooling\cite{zhu2019near}. Another difficulty arising in these processes is that the quasi-particles may be in nonequilibrium state due to driving from external bias.

In this work, we show that these processes can be conveniently analyzed by `bosonizing' a voltage-biased coherent electron conductor into bosonic reservoir with non-zero chemical potential. In the limit of weak electron-boson coupling, we obtain a Landauer formula to describe energy transport between electrons and bosons. This is possible since energy transport between electrons and bosons is always accompanied by the generation or annhilation of different kinds of electron-hole pairs (EHPs)\cite{headgordon_molecular_1995,dou2018perspective}. We thus generalizes the Landauer formalism to hybrid energy transport between possibly nonequilibrium baths, and provides a unified framework to understand energy transport in different thermal, thermoelectric and optoelectronic processes.

\section{Theory}
\subsection{System setup}
We consider a model system schematically shown in Fig.~\ref{fig:ehp} (a). The \emph{system} composed of an independent set of bosonic degrees of freedom (DOF) taken as a set of harmonic oscillators. It couples to two kinds of baths. One is an equilibrium boson bath (ph-bath), modeled by an infinite number of harmonic oscillators (bosonic modes). The other is an electron bath (e-bath), which itself includes a central part ($C$) and two electrodes ($L$ and $R$).  The e-bath may be driven into a nonequilibrium steady state by a voltage bias applied between the two electrodes. Without loss of generality, we assume that the system couples only to the central region of the e-bath. Energy transport between the two baths takes place through their simultaneous coupling to the system.
The electrons couple to the `displacement' of the system harmonic oscillators
\begin{equation}
H_{es} = \sum_{i,j,k} M^{k}_{ij}c^\dagger_i c_j u_k.
\label{eq:eboson}
\end{equation}
Here, $M^k_{ij}$ describes the coupling of the system mode $k$ to the electronic transition between states $i$ and $j$, and $u_k$ is the `displacement' operator of the system mode $k$. For phonons, it is the displacement, while for photons it is the vector potential. The system-ph-bath coupling is linear between harmonic oscillators and can be treated exactly.

\subsection{Electron-hole pair excitation}
Our key observation is that the energy transport between the system and the e-bath can be modeled by different kinds of reactions between EHPs in the e-bath and the bosonic modes in the system. The creation and annihilation of the bosonic mode is always accompanied by the recombination and creation of EHPs. These processes can be expressed in the form of reactions
\begin{align}
e_\alpha + h_\beta \rightleftharpoons b_n,
\label{eq:reaction}
\end{align}
where $e_\alpha$, $h_\beta$ and $b_n$ represent electron in electrode $\alpha$, hole in electrode $\beta$ and bosonic mode $b_n$ in the system. Equivalently, we can write
\begin{align}
e_\alpha \rightleftharpoons e_\beta + b_n,
\label{eq:reaction2}
\end{align}
representing inelastic electronic transition from electrode $\alpha$ to $\beta$, accompanied by emission of bosonic mode $n$ (forward process). The backward direction corresponds to absorption process.

\begin{figure}
	\centering
	\subfigure[]{
	\includegraphics[width=0.2\textwidth,angle=0]{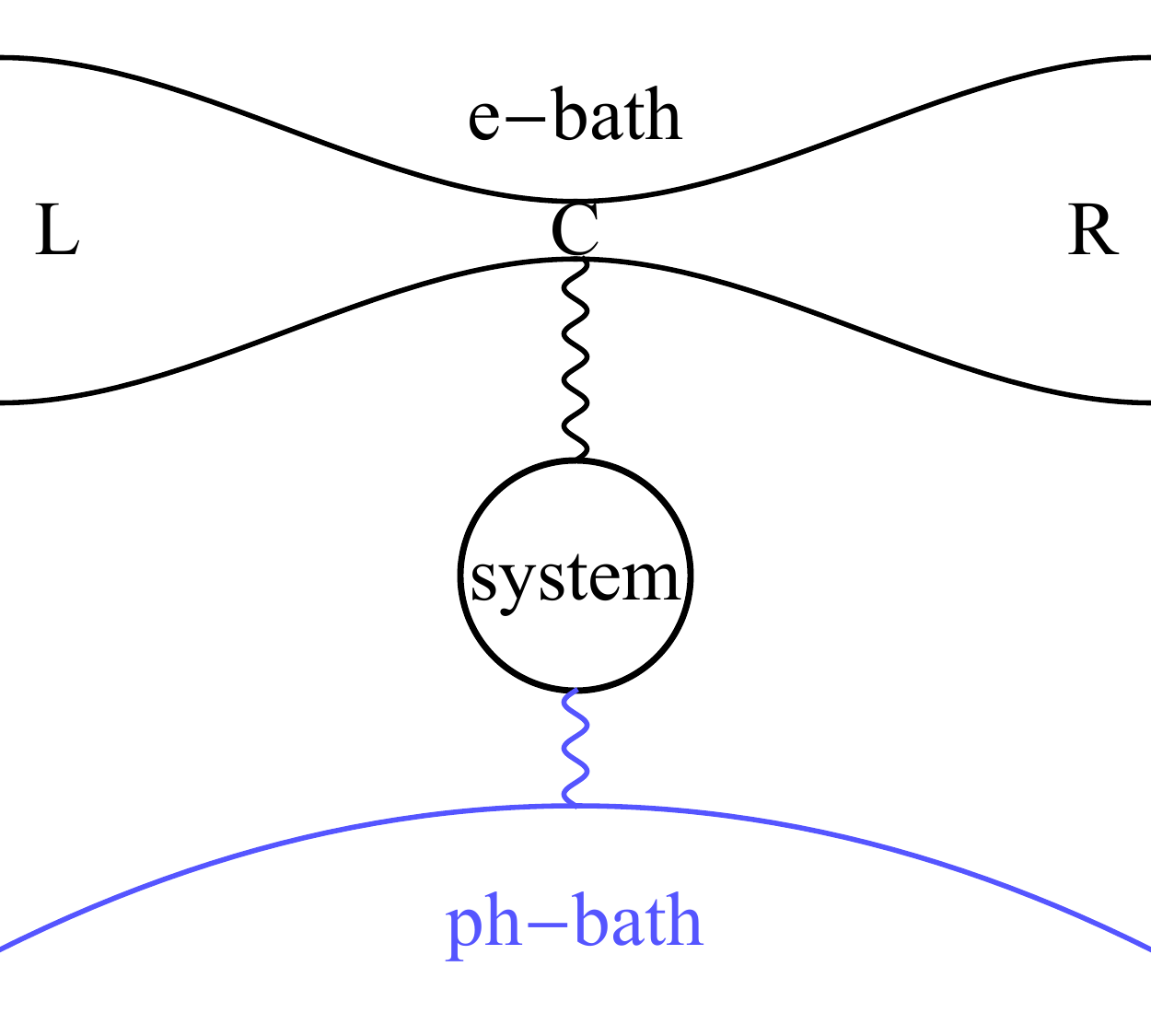}
	}
	\quad
	\subfigure[]{
	\includegraphics[width=0.2\textwidth,angle=0]{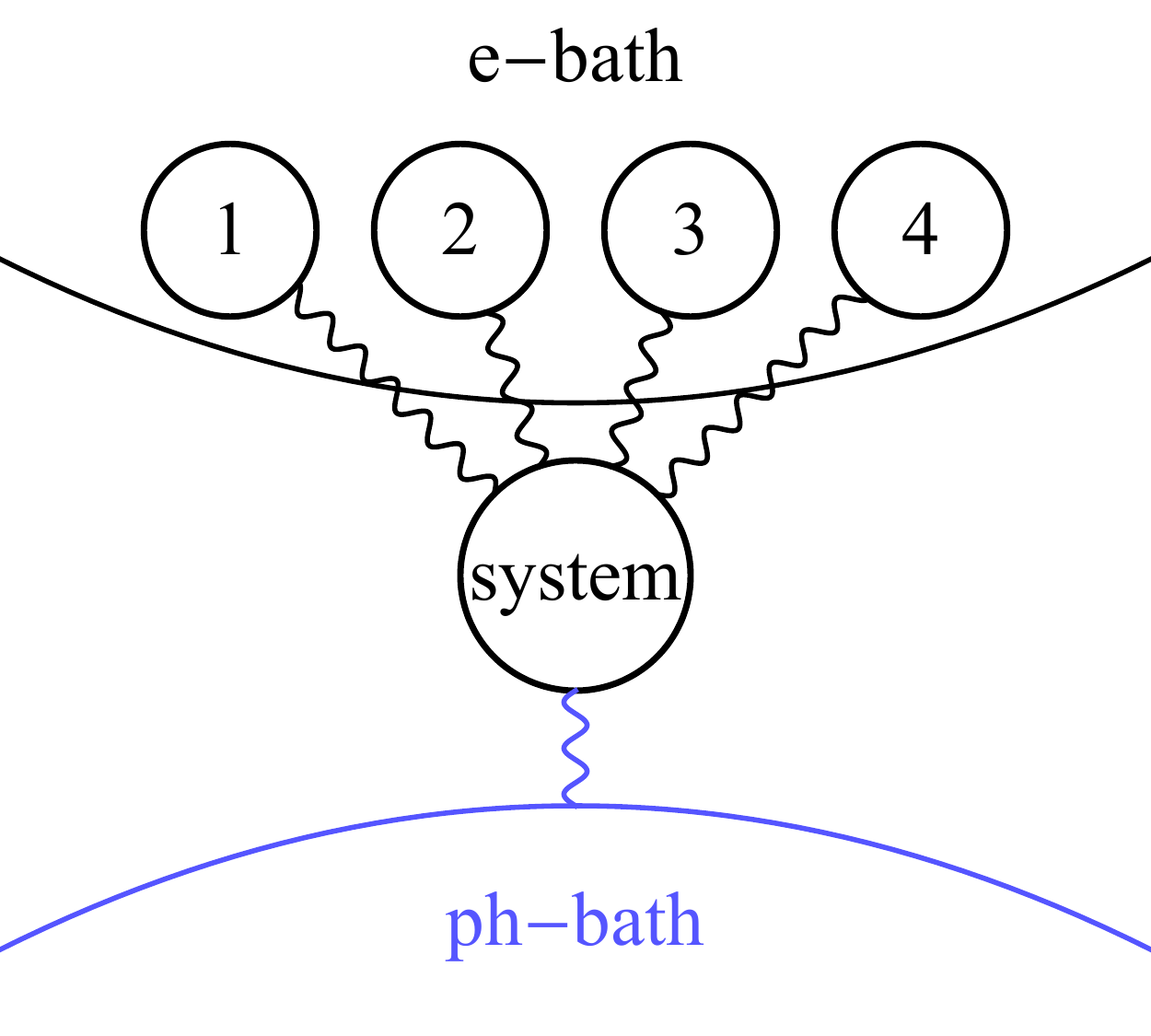}
	}
	\quad
	\subfigure[]{
	\includegraphics[width=0.2\textwidth,angle=0]{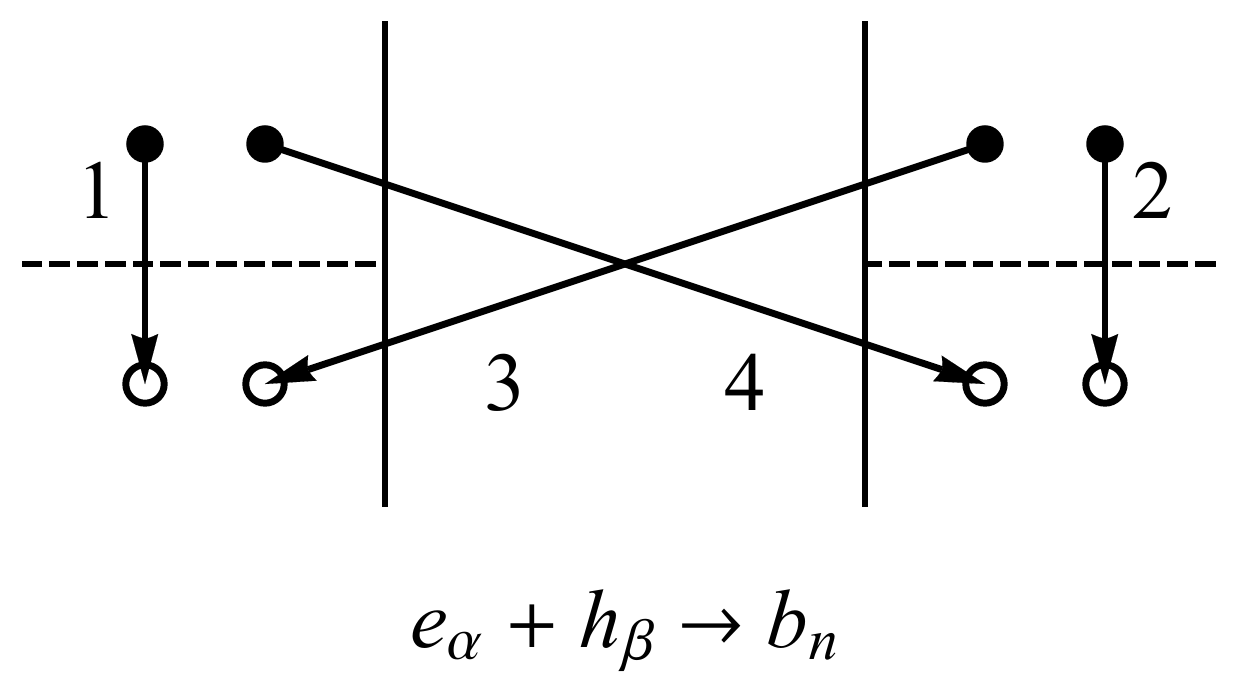}
	}
	\quad
	\subfigure[]{
	\includegraphics[width=0.2\textwidth,angle=0]{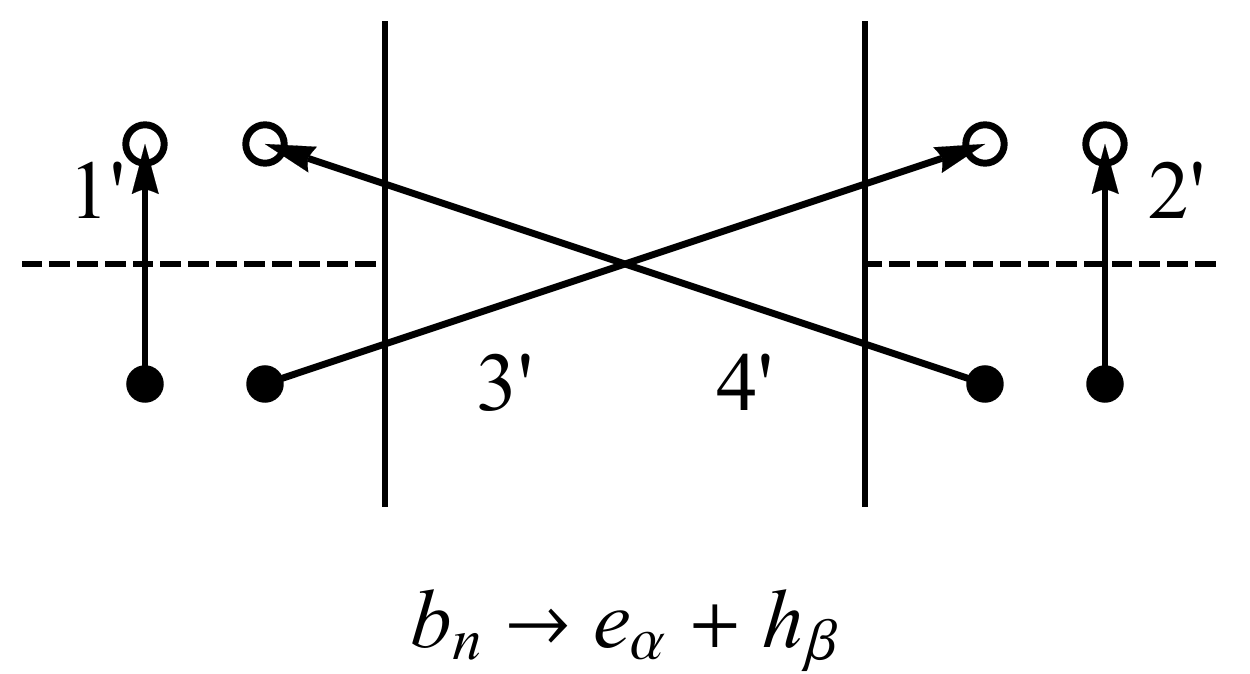}
	}
	\caption{(a) Schematics of the model we consider. The system consists a set of independent bosonic modes. It couples to an electron bath (e-bath), which is modeled as a conductor including a left (L) and a right (R) electrode, with temperature $T_{\rm e}$ and chemical potential $\mu_L$ and $\mu_R$, respectively. The system further couples to an external thermal bath (ph-bath) at temperature $T_{\rm {ph}}$. (b) The electron bath can be treated as four different kinds of electron-hole pair (EHP) baths (1-4), shown in (c). (c-d) Four kinds of EHP recombination (c) and excitation (d) processes. The EHPs are classified according to the spatial location of the electron ($e_\alpha$) and the hole ($h_\beta$).}
	\label{fig:ehp}
\end{figure}

There are four types of EHPs which we label by the spatial location of the electron ($\alpha$) and hole  ($\beta$) states. They are schematically shown in Fig.~\ref{fig:ehp} (c) and (d) for recombination and creation processes, respectively. They are denoted by EHP-$i$ ($i=1,2,3,4$) and are further divided into two groups. The intra-electrode type includes $1/LL$ and $2/RR$, and inter-electrode type includes $3/RL$, $4/LR$.  Additional to energy transfer between e-bath and system, the generation and recombination of inter-electrode EHPs also involves charge transport across the system. We take the energy of mode  and the EHPs to be positive.

A generalized detailed balance relation applies to each of reactions
\begin{align}
\frac{\tau _{\alpha\rightharpoonup\beta}}{\tau_{\alpha\leftharpoondown\beta}} = {\rm exp}\left[-\beta_{\rm B}(\hbar\Omega-\mu_{\alpha\beta})\right].
\label{eq:db}
\end{align}
Here, $\tau_{\alpha\rightharpoonup\beta}$ and $\tau_{\alpha\leftharpoondown\beta}$ are the reaction rates for the forward (boson emission) and backward (boson absorption) processes in Eq.~(\ref{eq:reaction}), respectively. They are obtained from the Fermi golden rule
\begin{align}
\tau_{\alpha\rightharpoonup\beta} &= \frac{2\pi}{\hbar}\sum_{i\in\alpha,f\in\beta}|M^m_{ij}|^2  \delta(\varepsilon_i-\varepsilon_f-\hbar\Omega)  \nonumber\\
&\times n_{\rm F}(\varepsilon_i-\mu_\alpha)(1-n_{\rm F}(\varepsilon_f-\mu_\beta)).
\end{align}
Here, $n_{\rm {F/B}}(\varepsilon,T) = \left[{\rm exp}\left(\beta_{\rm B} \varepsilon\right)\pm 1\right]^{-1}$ is the Fermi-Dirac/Bose-Einstein distribution, with $\beta_{\rm B}=(k_{\rm B}T)^{-1}$,  $\mu_{\alpha\beta}=\mu_\alpha-\mu_\beta$, and $M^m_{ij}=\langle \psi_{i}(\varepsilon_i)|M^{}|\psi_{f}(\varepsilon_f)\rangle$ is the transition matrix element from initial state $i$ in electrode $\alpha$ to final state $f$ in electrode $\beta$. The reverse rate $\tau_{\alpha\leftharpoondown\beta}$ can be written similarly.
Thus, when reaching equilibrium with the EHP bath $\alpha\beta$, the bosonic mode follows a Bose-Einstein distribution at temperature $T_{\rm e}$ and chemical potential $\mu_{\alpha\beta}$. For intra-electrode processes, $\mu_{\alpha\beta}=0$, we have the normal detailed balance relation, while for inter-electrode processes $\mu_{\alpha\beta}$ is determined by the applied voltage bias. Thus, the bosonic mode may acquire a non-zero chemical potential in nonequilibrium. This is consistent with the equilibrium condition for reaction \ref{eq:reaction}.

The key quantity to describe the EHP baths is the coupling-weighted power spectrum. It can be written as
\begin{align}
\tilde{\Pi}_{mn}^{\alpha\beta}(\omega) &= \left[n_{\rm B}(\hbar\omega-\mu_{\alpha\beta},T_{\rm e})+\frac{1}{2}\right]\Lambda_{mn}^{\alpha\beta}(\omega).
\label{eq:ppehp}
\end{align}
We have introduced the coupling-weighted EHP density of states (DOS)\cite{lu_current-induced_2012,lu2016electron}
\begin{align}
\Lambda_{mn}^{\alpha\beta}(\omega) &= -\sum_{i\in\alpha,f\in\beta}M^{m}_{fi}M^n_{if}  \delta(\varepsilon_i-\varepsilon_f-\hbar\omega)\nonumber\\
&\times (n_{\rm F}(\varepsilon_\alpha-\mu_\alpha,T_\alpha)-n_{\rm F}(\varepsilon_\beta-\mu_\beta,T_\beta))\nonumber\\
&=-\int \frac{d\varepsilon}{2\pi} {\rm tr}[M^m A_\alpha(\varepsilon) M^n A_\beta(\varepsilon-\hbar\omega)] \nonumber\\
&\times (n_{\rm F}(\varepsilon-\mu_\alpha,T_\alpha)-n_{\rm F}(\varepsilon-\hbar\omega-\mu_\beta,T_\beta)),
\label{eq:gamma}
\end{align}
which also characterizes the system dissipation due to coupling to the e-bath\cite{lu_current-induced_2012}. Here, $A_\alpha$ is electrode spectrum functional.
Equation~(\ref{eq:ppehp}) follows a form of the fluctuation-dissipation relation for an equilibrium ph-bath, albeit with a possibly non-zero chemical potential  $\mu_{\alpha\beta}$. The intra-electrode EHPs (i=1,2) are always in equilibrium with $\mu_{\alpha\alpha}=0$ and temperature $T_{\rm e}$. But the two inter-electrode EHPs (i=3, 4) have opposite chemical potential $\mu_{RL}=-\mu_{LR}$. They are non-zero when there is a voltage bias applied.  To this end, we have shown that the nonequilibrium e-bath can be divided into four EHP baths with different chemical potentials.
This effective model is shown in Fig.~\ref{fig:ehp} (b).

\subsection{Steady state mode population}
The reaction~\ref{eq:reaction} suggests that, when reaching steady state, the bosonic mode inherits the chemical potential of the EHPs. Thus, the bosonic mode may acquire a non-zero chemical potential. This is best illustrated by performing a mode population analysis.

To simplify the analysis, we consider one bosonic mode with angular frequency $\Omega$. A simple master equation for the mode population $N$ can be established by considering the forward and backward reaction processes
\begin{align}
	\dot{N} = \sum_{\alpha\beta}\left[\tau_{\alpha\rightharpoonup\beta} (N+1) - \tau_{\alpha\leftharpoondown\beta} N\right].
\end{align}
The steady state population of mode is obtained by setting $\dot{N}=0$, which is written as
\begin{align}
	N = \frac{1}{\sum_{\alpha\beta}\tau_{\alpha\leftharpoondown\beta}/\sum_{\alpha\beta}\tau_{\alpha\rightharpoonup\beta}-1}.
\end{align}
In equilibrium ($\mu_L=\mu_R$), we obtain the standard Bose-Einstein distribution with temperature $T_{\rm e}$ and zero chemical potential. When there is voltage bias applied ($\mu_L\neq\mu_R$), the final distribution can not be written as a simple form. Normally, an effective temperature $T_{\rm eff}$ is defined by assuming $N$ follows the Bose-Einstein distribution with zero chemical potential
\begin{align}
	k_{\rm B}T_{\rm eff} = \frac{\hbar\Omega}{{\rm ln}(1+N^{-1})}.
\end{align}
According to previous discussion, we can equivalently define an effective chemical potential by assuming $N$ follows the Bose-Einstein distribution at $T_{\rm e}$
\begin{align}
	\mu_{\rm eff} = \hbar\Omega - k_{\rm B}T_{\rm e}{\rm ln}(1+N^{-1}).
\end{align}
These are two equivalent equivalent ways of characterizing the nonequilibrium steady state of the vibrational mode. The two effective parameters are related via
\begin{align}
	T_{\rm eff} = \frac{T_{\rm e}}{1-\mu_{\rm eff}/(\hbar\Omega)}.
	\label{eq:tmu}
\end{align}
Several comments are noteworthy at this point. Firstly, in the presence of voltage bias, if the emission process is enhanced more than the absorption process, we have a negative $\mu_{\rm eff}$ and consequently $T_{\rm eff}>T_{\rm e}$. The result is heating of the bosonic mode. In the limiting case shown in Fig.~\ref{fig:resonant}(a), resonant enhancement may lead to the extreme case of $\mu_{\rm eff}=\hbar\Omega$, or $T_{\rm eff}\to +\infty$. This marks the instability of the bosonic mode. This case has been analyzed in details in Ref.~\onlinecite{lu2011laserlike}. The instability means that the perturbative analysis is not applicable any more\cite{nitzan2018kinetic}. It can be avoided by introducing additional coupling to the ph-bath. The validity of $T_{\rm eff}$ in this case will be analyzed elsewhere\cite{Wang-preprint}. In the other limiting case (Fig.~\ref{fig:resonant}(b)), the absorption process is resonantly enhanced, resulting in $\mu_{\rm eff}>0$ or $T_{\rm eff}<T_{{\rm e}}$. In this regime, the voltage bias is used to cool the bosonic mode below $T_{\rm e}$.

\subsection{Energy transport}
Within this effective EHP model, hybrid energy transport between  electrons and system bosons can be treated as bosonic transport.
To the lowest order approximation, we arrive at a  Landauer-like formula for the energy and particle transport from e-bath to the system as a summation of contributions from all the EHP baths
\begin{align}
J &= \sum_{\alpha,\beta}\int_0^{+\infty}\frac{d\omega}{2\pi}\hbar\omega\ {\rm Tr}[\Lambda^{\alpha\beta}(\omega)\mathcal{A}_{\rm {ph}}(\omega)]\nonumber\\
&\times [n_{\rm B}(\omega-\mu_{\alpha\beta},{T}_{{\rm e}})-n_{\rm B}(\omega,T_{\rm {ph}})].
\label{eq:jjtrans}
\end{align}
Here, $J$ is the energy flux from e-bath to ph-bath. $T_{\rm e}$ and $T_{\rm ph}$ are the temperature of the e-bath and ph-bath, respectively. The trace Tr is over system DOF, with $\mathcal{A}_{\rm ph}$ the spectral function of the system due to coupling to the ph-bath. We can write it in terms of the non-interacting boson Green's function $D^{r/a}$ and self-energy $\Pi_{\rm ph}^{r/a}$ as $\mathcal{A}_{\rm {ph}}=iD^r(\Pi_{\rm ph}^r-\Pi_{\rm ph}^a)D^a$\cite{lu2016electron}. The summation over $\alpha\beta$ includes contributions from all the four types of EHPs. Each of them contributes to one transport channel.

\begin{figure}[h]
	
	\centering
	\subfigure[]{
		\includegraphics[width=0.2\textwidth,angle=0]{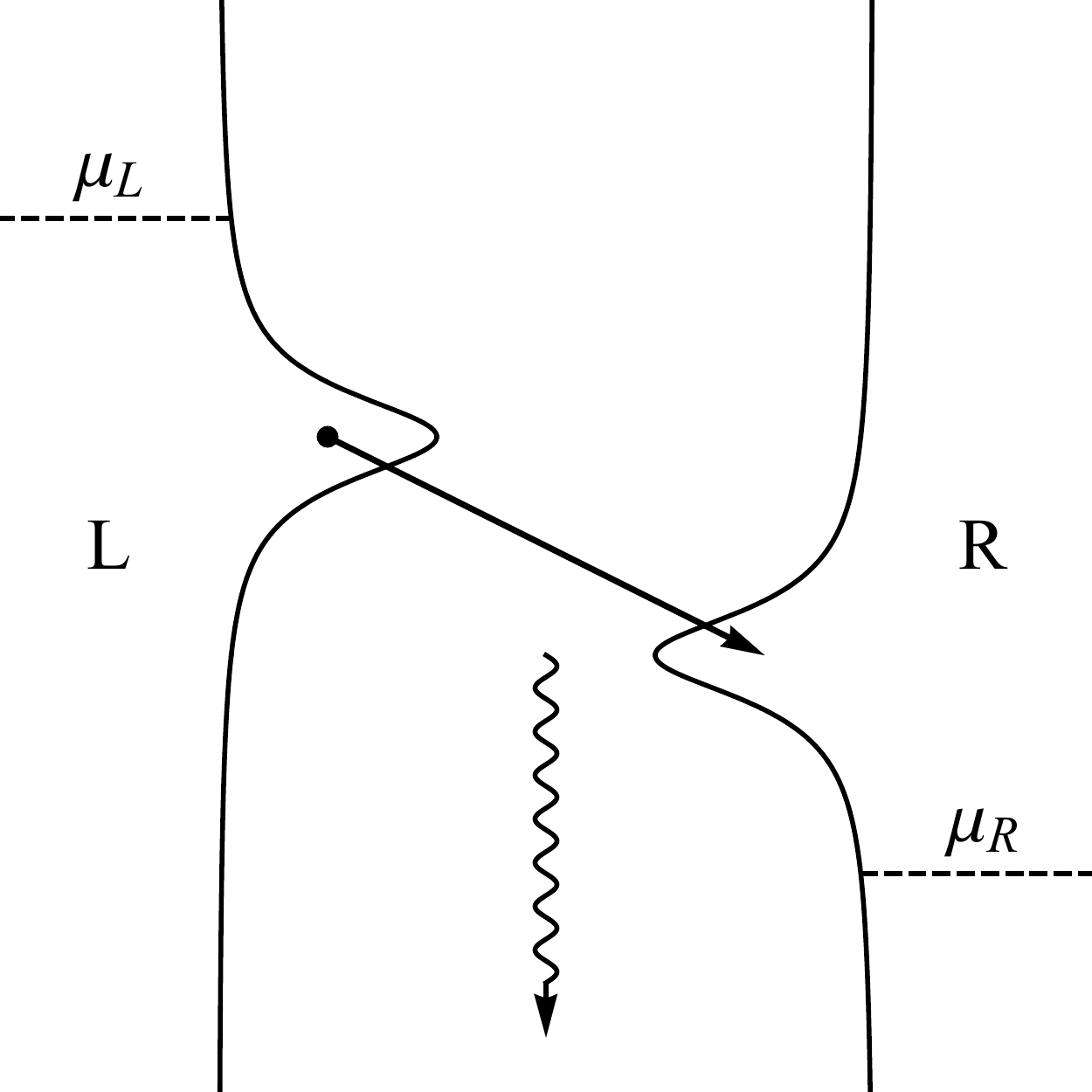}
		
	}
	\quad
	\subfigure[]{
		\includegraphics[width=0.2\textwidth,angle=0]{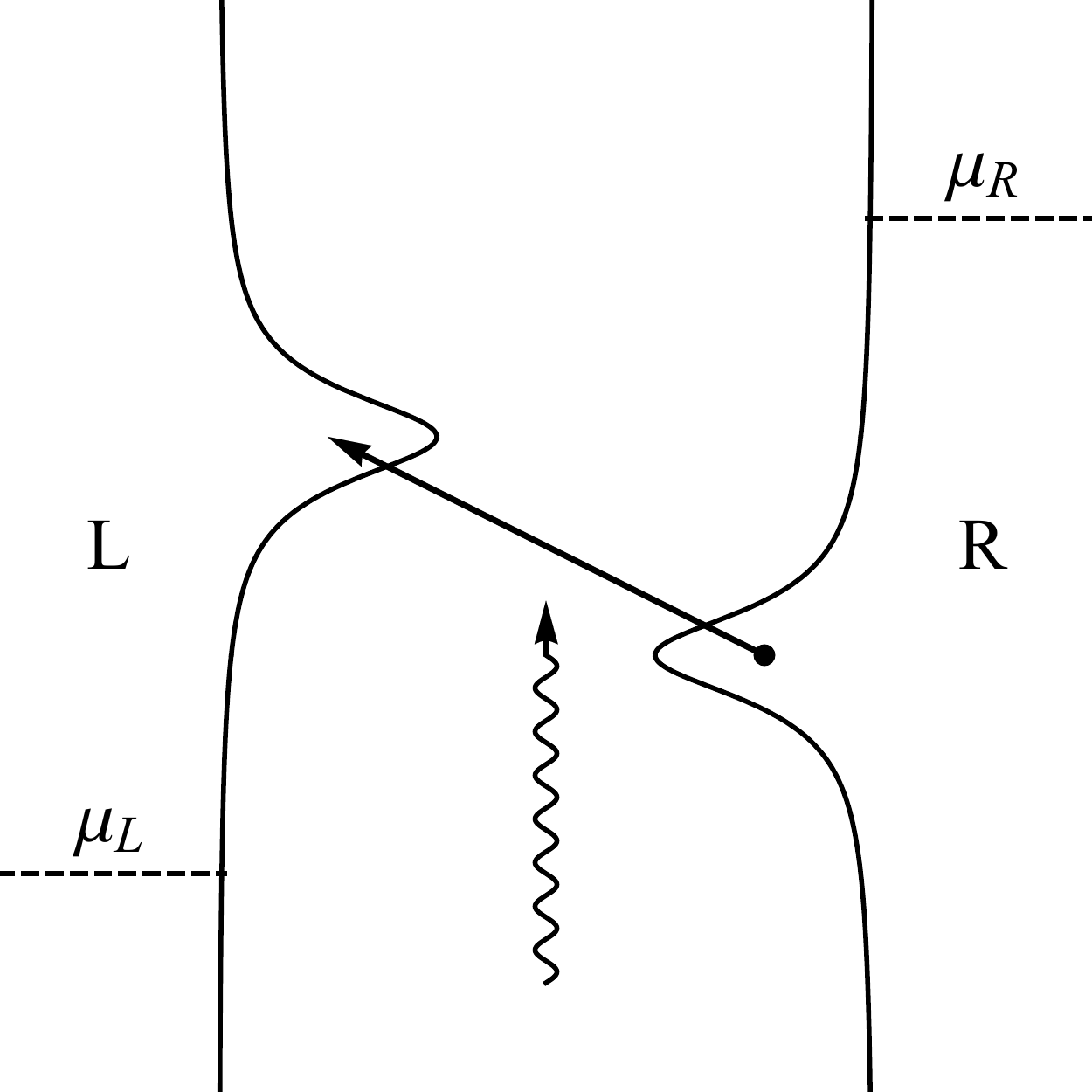}
	}
	\caption{Two limiting cases of nonequilibrium reservoir engineering. In (a), we have a filled electronic level $\varepsilon_L$ that couples to the left electrode with chemical potential $\mu_L$, and an empty level $\varepsilon_R$ that couples to the right electrode with chemical potential $\mu_R$. We have $\mu_L>\mu_R$. Heating of the bosonic mode is due to resonant recombination of inter-electrode EHPs (process 4 in Fig.~\ref{fig:ehp}). In (b), the situation is reversed. The left state $\varepsilon_L$ is empty, while the right state $\varepsilon_R$ is filled. When $\mu_R>\mu_L$, the e-bath can be used to cool the bosonic mode through creation of inter-electrode EHPs (process $4'$ in Fig.~\ref{fig:ehp}). }
	\label{fig:resonant}
\end{figure}

In the following we show several applications of this central result.
To be more specific, we consider a minimum model of the e-bath shown in Fig.~\ref{fig:resonant}. We have two electronic states $1$ and $2$ (on-site energies $\varepsilon_1$ and $\varepsilon_2$) couple to the electrodes $L$ and $R$ with coupling parameter $\gamma_1$ and $\gamma_2$, respectively.
Electron hopping between the two states is assisted by one bosonic mode, which at the same time couples to a ph-bath with coupling constant $\gamma_{\rm {ph}}$.

\section{Applications}
\subsection{Non-reciprocal heat transport}
Firstly, we consider the situation where the e-bath and ph-bath are in their own thermal equilibrium at two different temperature $T_{\rm e}$ and $T_{\rm {ph}}$. This indicates that  $\mu_{L}=\mu_{R}$ and $T_{L}=T_R=T_{\rm e}$.
If we ignore the energy dependence of $A$ in Eq.~(\ref{eq:gamma}),
$\Lambda_{mn}(\omega)=\hbar\omega{\rm tr}[M^m A M^n A]$
with $A=A_L+A_R$. Consequently,  the transmission $\mathcal{T}={\rm Tr}[\Lambda\mathcal{A}_{\rm {ph}}]$ does not depend on $T_{\rm e}$. Equation~(\ref{eq:jjtrans}) reduces to the Landauer formula for heat transport between two harmonic thermal baths. Thus, the EHPs behave as linear harmonic oscillator thermal baths.

\begin{figure}[]
	\includegraphics[width=0.45\textwidth,angle=0]{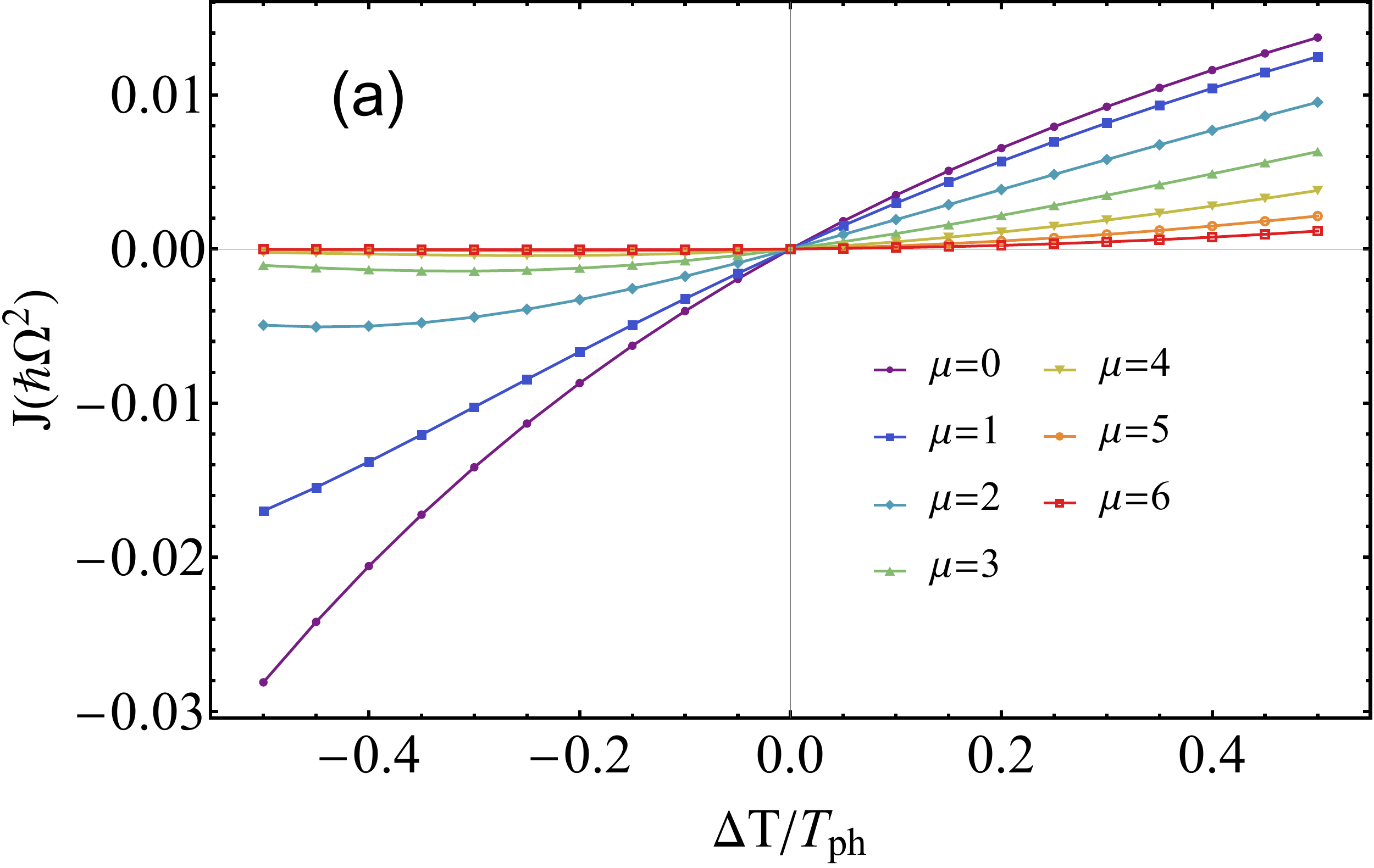}
	\includegraphics[width=0.44\textwidth,angle=0]{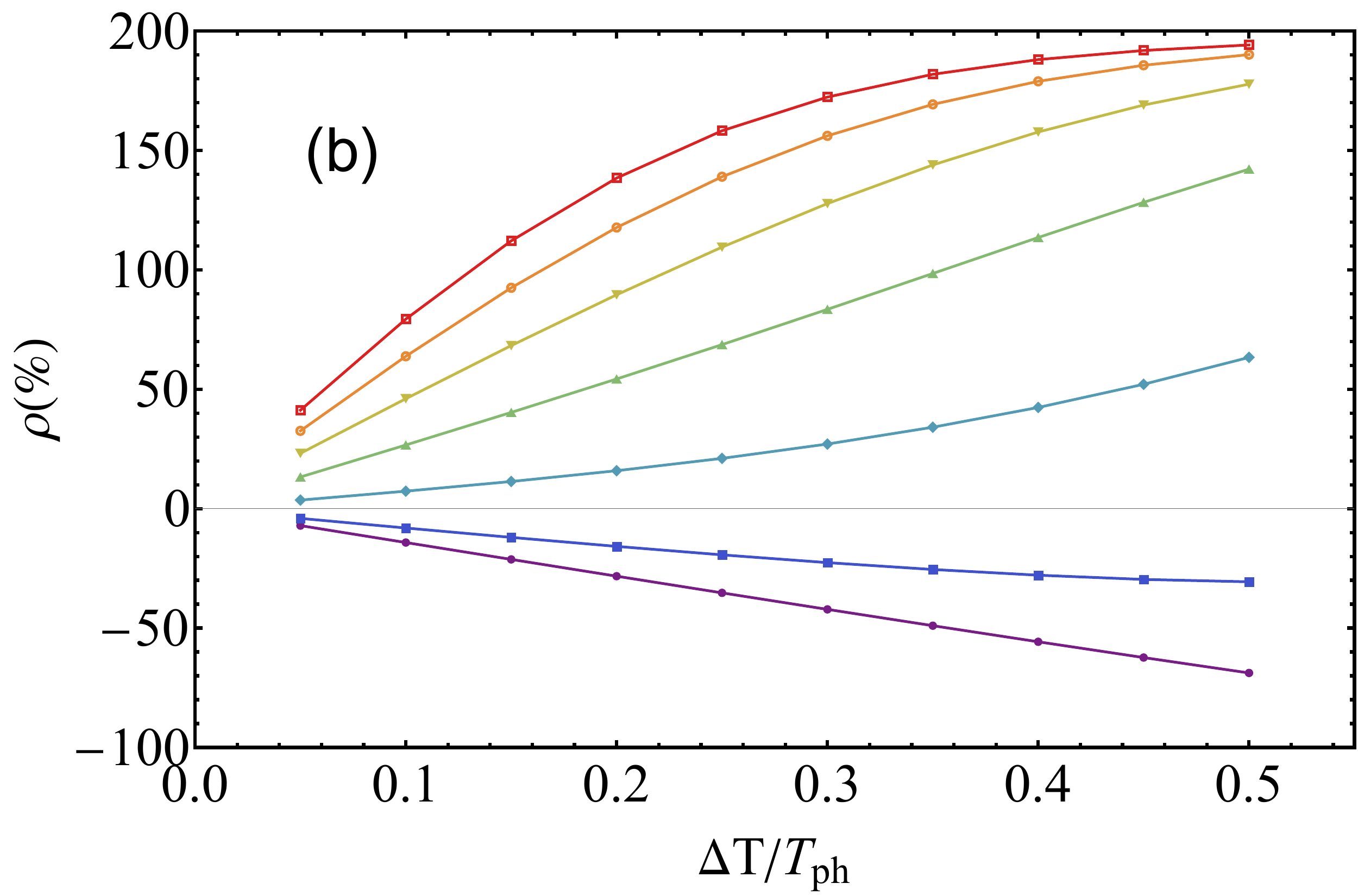}	
	
	\caption{Non-reciprocal heat transport in a  double dot junction shown in Fig.~\ref{fig:resonant} (a). (a) Heat current as a function of temperature difference $\Delta T/T_{\rm ph}$ for different chemical potentials. (b) Rectification ratio $\rho=(|J(|\Delta T|)|-|J(-|\Delta T|)|)/(|J(|\Delta T|)|+|J(-|\Delta T|)|)/2$ as a function of $\Delta T/T_{\rm ph}$ for different chemical potentials. We consider only one bosonic mode, whose energy is taken as unit energy. The following parameters are used in the calculation: $\varepsilon_L=0.5$, $\varepsilon_R=-0.5$, $\gamma_L=\gamma_R=0.5,m=0.5$, $\hbar\Omega=1$, $k_{\rm B}=1$. }
	\label{fig:JT}
	
\end{figure}

On the other hand, if we consider the energy dependence of $A(\varepsilon)$, $\Gamma(\omega)$, $\mathcal{T}^{}$ will depend on $T_{\rm e}$. Energy transport becomes anharmonic. In this case, non-reciprocal energy flow is possible, i.e., $J(\Delta T)\neq J(-\Delta T)$, with $\Delta T=T_{\rm e}-T_{\rm {ph}}$. We thus find a necessary condition for non-reciprocal energy transport in a hybrid electron-boson system: the electron DOS in the thermal window near the chemical potential has to be energy dependent\cite{zhang2013thermal,ren2013heat}. For normal metal electrode, the energy scale of electrons is much larger than the thermal energy, leading to a flat DOS. The energy dependence of $A(\varepsilon)$ can be engineered by changing the electronic states of the central part. For example, discrete energy levels of a molecular junction or quantum dot can be used.
In Fig.~\ref{fig:JT} we have considered a two-dot junction shown in Fig.~\ref{fig:resonant} (a). We set $\mu_L=\mu_R=\mu$ and $T_{\rm e} \neq T_{\rm {ph}}$ to consider heat transport. The electronic DOS shows an energy dependent Lorentzian shape. This gives rise to non-reciprocal heat transport between e-bath and ph-bath.

\begin{figure}[h]
	\includegraphics[width=0.45\textwidth,angle=0]{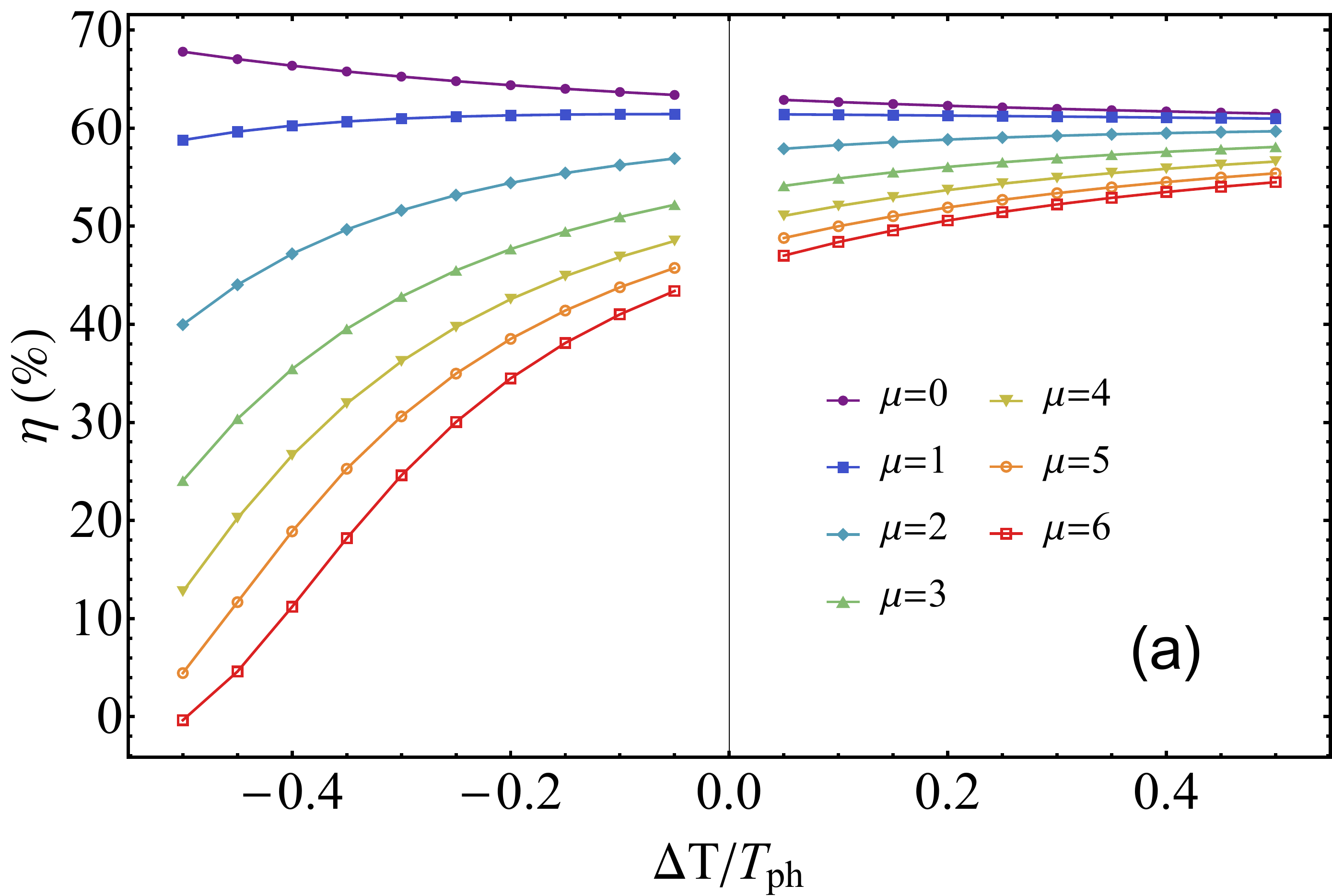}
	\includegraphics[width=0.48\textwidth,angle=0]{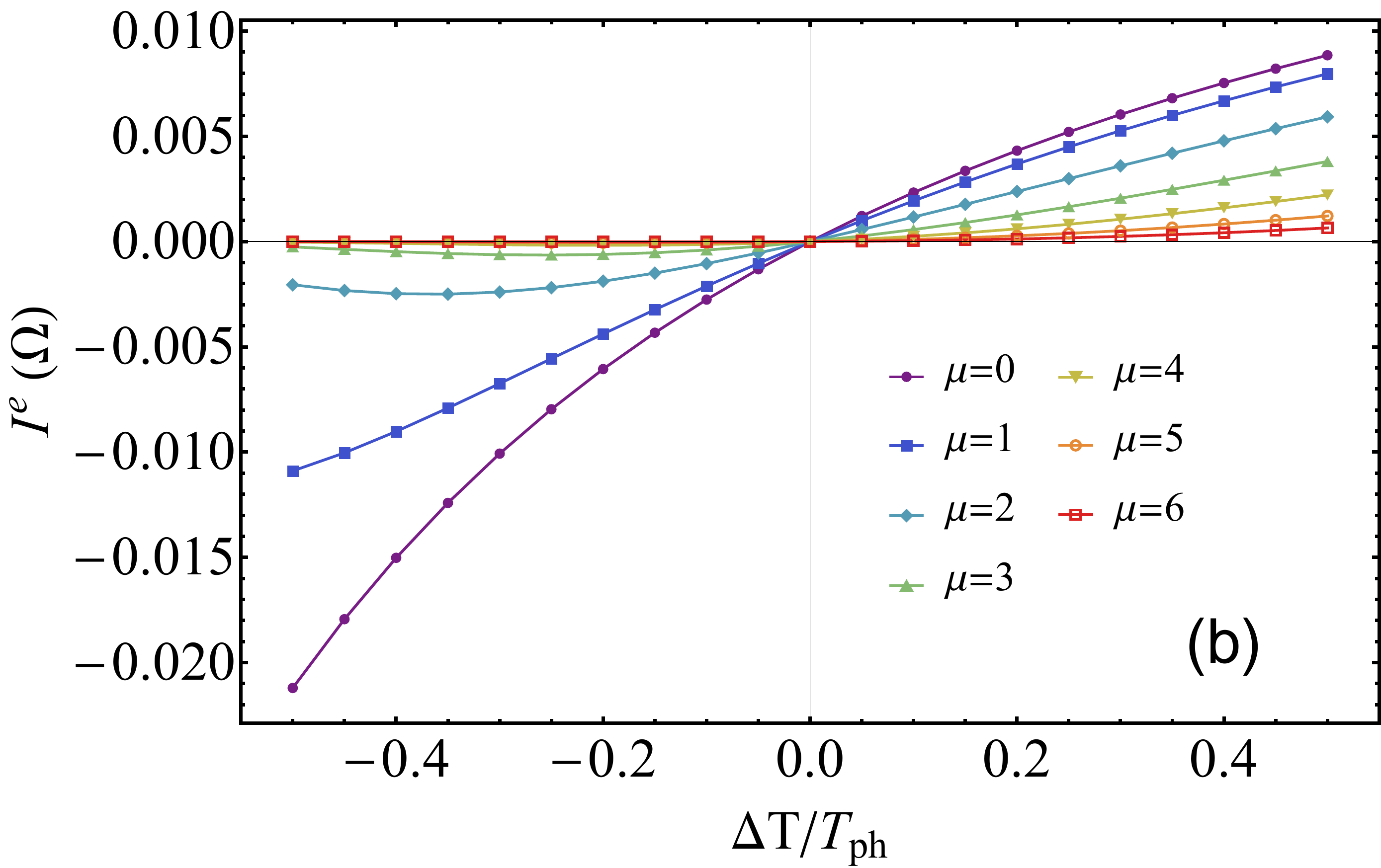}	
	
	\caption{Thermoelectric efficiency $\eta$ (a)  and electron particle flux $I^e$ (b) as a function of temperature difference between e-bath and ph-bath $\Delta T/T_{\rm {ph}}$. The parameters are the same as Fig.~\ref{fig:JT}.   }
	\label{fig:coldspot}
\end{figure}

\subsection{Hybrid thermoelectric transport}
 We can also study the thermoelectric transport of the temperature-biased electron-boson junction. When $T_{\rm {ph}}\neq T_{\rm e}$, in addition to the heat transport between system and e-bath, an electrical current may also be induced between the two electrodes\cite{entinwuhlman2010three,sanchez2011optimal}. In our EHP picture, this is realized through coupling of the bosonic mode with two inter-electrode EHPs. Since they contribute to the electrical current with opposite directions, in order to get a non-zero electrical current, these two channels should not get canceled. We can write the electron particle flux as
 \begin{align}
 I^e &= \sum_{\alpha,\beta} (\delta_{\alpha L}\delta_{\beta R} - \delta_{\alpha R}\delta_{\beta L})   \int_0^{+\infty}\frac{d\omega}{2\pi}\ {\rm Tr}[\Lambda^{\alpha\beta}(\omega)\mathcal{A}_{\rm {ph}}(\omega)]\nonumber\\
 &\times [n_{\rm B}(\omega-\mu_{\alpha\beta},{T}_{{\rm e}})-n_{\rm B}(\omega,T_{\rm {ph}})].
 \label{eq:ietrans}
 \end{align}
 Here, $\delta_{\alpha/\beta, L/R}$  are the Kronecker delta functions. For simplicity, we introduce thermoelectric efficiency $\eta$ as the ratio between electron particle flux and phonon particle flux $\eta= I^e / (J^{\rm {ph}}/\hbar\Omega)$.

 The resonant situation in Fig.~\ref{fig:resonant} can be used to enhance one of the two channels. In Fig.~\ref{fig:coldspot} (b), we show the thermoelectric current induced by the temperature different $\Delta T$ for different chemical potentials $\mu_L=\mu_R$ in the case of Fig.~\ref{fig:resonant} (a). The efficiency $\eta$ in Fig.~\ref{fig:coldspot} (a) is the largest when the chemical potential is in between $\varepsilon_L$ and $\varepsilon_R$, where the resonant enhancement is the most prominent.
 Previously, electrical current generated from a phonon hot-spot ($T_{\rm {ph}}>T_{\rm e}$) has been considered\cite{entinwuhlman2010three}. Our results show that the opposite ($T_{\rm {ph}}<T_{\rm e}$) is also possible, where electricity is generated by cooling the ph-bath. This demonstrates the decoupling of heat and charge transport as an advantage of thermoelectricity in hybrid nano-junctions.

\subsection{Electronic cooling of bosonic mode}
We now turn on the voltage bias in the e-bath.
The applied voltage changes the initial and final electron states of the EHP excitation. Thus, the EHP DOS can be modified by voltage. More importantly, the inter-electrode EHPs acquire a non-zero chemical potential, EHP-4 has a chemical potential of $\mu_{LR}$, while EHP-3 gets a chemical potential with opposite value $\mu_{RL}$.
\begin{figure}[h]
	
	\includegraphics[width=0.5\textwidth,angle=0]{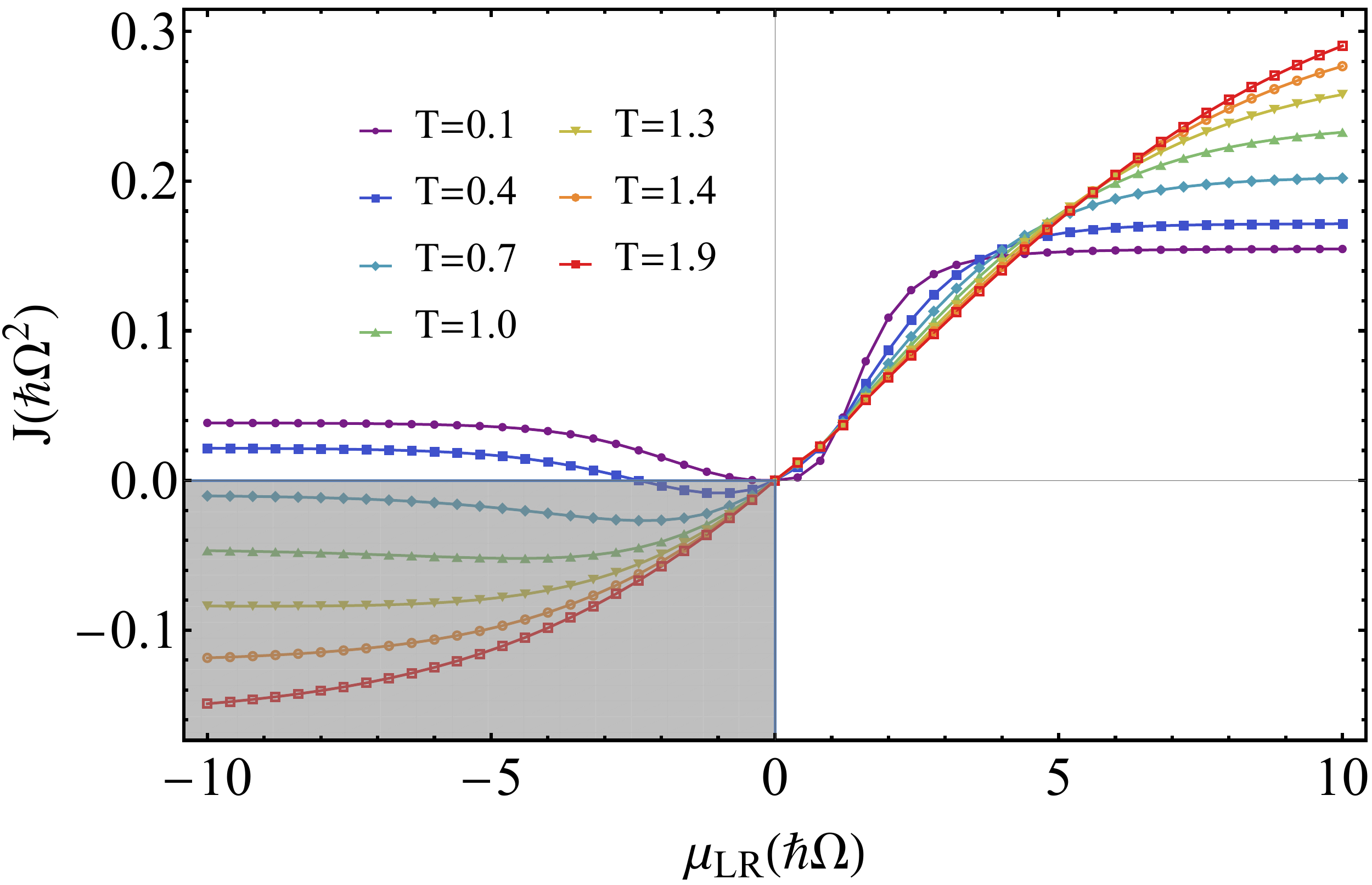}	
	\caption{Energy current $J$ from the e-bath to the bosonic mode as a function of chemical potential $\mu_{LR}$, corresponding to the situation in Fig.~\ref{fig:resonant} (b). Negative $J$ (gray shaded area) means cooling of the bosonic mode.  }
	\label{fig:cooling}
\end{figure}
Change of the chemical potential breaks the equilibrium in the reaction, and drives the energy transport between e-bath and the system.
Direction of energy flow depends on the relative magnitude of two fluxes. It can be engineered by tuning the electronic band structure, or more specifically, the transition probability of the two types of EHP excitation.

In the case shown in Fig.~\ref{fig:resonant} (b), process $4'$ is resonantly enhanced. Electronic cooling becomes possible using this resonant enhancement. This is demonstrated in Fig.~\ref{fig:cooling}, where the heat current from the e-bath to the system $J$ is plotted as a function of voltage bias $\mu_{LR}$ while keeping temperature fixed $T_{\rm e}=T_{\rm {ph}}=T$. For negative bias, we observe a negative $J$ regime. The range of this regime gets larger for higher temperature $T$. This is the electronic cooling of the bosonic mode. Very recently, experimental demonstration of near field radiative cooling using a reversely biased $p$-$n$ junction has been demonstrated \cite{zhu2019near}. The experimental results can be understood using this simple model.

\section{Conclusions}
In summary, we have shown that a normal two-probe electron conductor can be effectively viewed as EHP baths with chemical potential determined by the applied voltage bias. This is made possible by introducing the inter-electrode charge transfer EHPs. Properties of the EHP baths can be engineered through tuning the parameters of the conductor and the external voltage bias. This bath engineering provides an efficient way of controlling hybrid energy and thermoelectric transport in electron-boson junctions.

\begin{acknowledgements}
The authors thank J.-S. Wang and M. Brandbyge for discussions. This work is supported by the National Natural Science Foundation of China
(Grant No. 21873033), the National Key Research and Development Program of China
(Grant No. 2017YFA0403501) and the program for HUST academic frontier youth team.
\end{acknowledgements}

\bibliography{LB,gle-review}

\begin{thebibliography}{41}%
\makeatletter
\providecommand \@ifxundefined [1]{%
 \@ifx{#1\undefined}
}%
\providecommand \@ifnum [1]{%
 \ifnum #1\expandafter \@firstoftwo
 \else \expandafter \@secondoftwo
 \fi
}%
\providecommand \@ifx [1]{%
 \ifx #1\expandafter \@firstoftwo
 \else \expandafter \@secondoftwo
 \fi
}%
\providecommand \natexlab [1]{#1}%
\providecommand \enquote  [1]{``#1''}%
\providecommand \bibnamefont  [1]{#1}%
\providecommand \bibfnamefont [1]{#1}%
\providecommand \citenamefont [1]{#1}%
\providecommand \href@noop [0]{\@secondoftwo}%
\providecommand \href [0]{\begingroup \@sanitize@url \@href}%
\providecommand \@href[1]{\@@startlink{#1}\@@href}%
\providecommand \@@href[1]{\endgroup#1\@@endlink}%
\providecommand \@sanitize@url [0]{\catcode `\\12\catcode `\$12\catcode
  `\&12\catcode `\#12\catcode `\^12\catcode `\_12\catcode `\%12\relax}%
\providecommand \@@startlink[1]{}%
\providecommand \@@endlink[0]{}%
\providecommand \url  [0]{\begingroup\@sanitize@url \@url }%
\providecommand \@url [1]{\endgroup\@href {#1}{\urlprefix }}%
\providecommand \urlprefix  [0]{URL }%
\providecommand \Eprint [0]{\href }%
\providecommand \doibase [0]{http://dx.doi.org/}%
\providecommand \selectlanguage [0]{\@gobble}%
\providecommand \bibinfo  [0]{\@secondoftwo}%
\providecommand \bibfield  [0]{\@secondoftwo}%
\providecommand \translation [1]{[#1]}%
\providecommand \BibitemOpen [0]{}%
\providecommand \bibitemStop [0]{}%
\providecommand \bibitemNoStop [0]{.\EOS\space}%
\providecommand \EOS [0]{\spacefactor3000\relax}%
\providecommand \BibitemShut  [1]{\csname bibitem#1\endcsname}%
\let\auto@bib@innerbib\@empty
%</preamble>
\bibitem [{\citenamefont {Imry}\ and\ \citenamefont
  {Landauer}(1999)}]{imry1999conductance}%
  \BibitemOpen
  \bibfield  {author} {\bibinfo {author} {\bibfnamefont {Y.}~\bibnamefont
  {Imry}}\ and\ \bibinfo {author} {\bibfnamefont {R.}~\bibnamefont
  {Landauer}},\ }\href {\doibase 10.1103/RevModPhys.71.S306} {\bibfield
  {journal} {\bibinfo  {journal} {Rev. Mod. Phys.}\ }\textbf {\bibinfo {volume}
  {71}},\ \bibinfo {pages} {S306} (\bibinfo {year} {1999})}\BibitemShut
  {NoStop}%
\bibitem [{\citenamefont {Ojanen}\ and\ \citenamefont
  {Jauho}(2008)}]{ojanen2008mesoscopic}%
  \BibitemOpen
  \bibfield  {author} {\bibinfo {author} {\bibfnamefont {T.}~\bibnamefont
  {Ojanen}}\ and\ \bibinfo {author} {\bibfnamefont {A.-P.}\ \bibnamefont
  {Jauho}},\ }\href {\doibase 10.1103/PhysRevLett.100.155902} {\bibfield
  {journal} {\bibinfo  {journal} {Phys. Rev. Lett.}\ }\textbf {\bibinfo
  {volume} {100}},\ \bibinfo {pages} {155902} (\bibinfo {year}
  {2008})}\BibitemShut {NoStop}%
\bibitem [{\citenamefont {Biehs}\ \emph {et~al.}(2010)\citenamefont {Biehs},
  \citenamefont {Rousseau},\ and\ \citenamefont
  {Greffet}}]{biehs2010mesoscopic}%
  \BibitemOpen
  \bibfield  {author} {\bibinfo {author} {\bibfnamefont {S.-A.}\ \bibnamefont
  {Biehs}}, \bibinfo {author} {\bibfnamefont {E.}~\bibnamefont {Rousseau}}, \
  and\ \bibinfo {author} {\bibfnamefont {J.-J.}\ \bibnamefont {Greffet}},\
  }\href {\doibase 10.1103/PhysRevLett.105.234301} {\bibfield  {journal}
  {\bibinfo  {journal} {Phys. Rev. Lett.}\ }\textbf {\bibinfo {volume} {105}},\
  \bibinfo {pages} {234301} (\bibinfo {year} {2010})}\BibitemShut {NoStop}%
\bibitem [{\citenamefont {Zhang}\ \emph {et~al.}(2018)\citenamefont {Zhang},
  \citenamefont {L{\"u}},\ and\ \citenamefont {Wang}}]{zhang2018energy}%
  \BibitemOpen
  \bibfield  {author} {\bibinfo {author} {\bibfnamefont {Z.-Q.}\ \bibnamefont
  {Zhang}}, \bibinfo {author} {\bibfnamefont {J.-T.}\ \bibnamefont {L{\"u}}}, \
  and\ \bibinfo {author} {\bibfnamefont {J.-S.}\ \bibnamefont {Wang}},\ }\href
  {\doibase 10.1103/PhysRevB.97.195450} {\bibfield  {journal} {\bibinfo
  {journal} {Phys. Rev. B}\ }\textbf {\bibinfo {volume} {97}},\ \bibinfo
  {pages} {195450} (\bibinfo {year} {2018})}\BibitemShut {NoStop}%
\bibitem [{\citenamefont {Ben-Abdallah}\ and\ \citenamefont
  {Biehs}(2014)}]{benabdallah2014near}%
  \BibitemOpen
  \bibfield  {author} {\bibinfo {author} {\bibfnamefont {P.}~\bibnamefont
  {Ben-Abdallah}}\ and\ \bibinfo {author} {\bibfnamefont {S.-A.}\ \bibnamefont
  {Biehs}},\ }\href {\doibase 10.1103/PhysRevLett.112.044301} {\bibfield
  {journal} {\bibinfo  {journal} {Phys. Rev. Lett.}\ }\textbf {\bibinfo
  {volume} {112}},\ \bibinfo {pages} {044301} (\bibinfo {year}
  {2014})}\BibitemShut {NoStop}%
\bibitem [{\citenamefont {Rego}\ and\ \citenamefont
  {Kirczenow}(1998)}]{rego1998quantized}%
  \BibitemOpen
  \bibfield  {author} {\bibinfo {author} {\bibfnamefont {L.~G.~C.}\
  \bibnamefont {Rego}}\ and\ \bibinfo {author} {\bibfnamefont {G.}~\bibnamefont
  {Kirczenow}},\ }\href {\doibase 10.1103/PhysRevLett.81.232} {\bibfield
  {journal} {\bibinfo  {journal} {Phys. Rev. Lett.}\ }\textbf {\bibinfo
  {volume} {81}},\ \bibinfo {pages} {232} (\bibinfo {year} {1998})}\BibitemShut
  {NoStop}%
\bibitem [{\citenamefont {Mingo}\ and\ \citenamefont
  {Broido}(2005)}]{mingo2005carbon}%
  \BibitemOpen
  \bibfield  {author} {\bibinfo {author} {\bibfnamefont {N.}~\bibnamefont
  {Mingo}}\ and\ \bibinfo {author} {\bibfnamefont {D.~A.}\ \bibnamefont
  {Broido}},\ }\href {\doibase 10.1103/PhysRevLett.95.096105} {\bibfield
  {journal} {\bibinfo  {journal} {Phys. Rev. Lett.}\ }\textbf {\bibinfo
  {volume} {95}},\ \bibinfo {pages} {096105} (\bibinfo {year}
  {2005})}\BibitemShut {NoStop}%
\bibitem [{\citenamefont {Yamamoto}\ and\ \citenamefont
  {Watanabe}(2006)}]{yamamoto2006nonequilibrium}%
  \BibitemOpen
  \bibfield  {author} {\bibinfo {author} {\bibfnamefont {T.}~\bibnamefont
  {Yamamoto}}\ and\ \bibinfo {author} {\bibfnamefont {K.}~\bibnamefont
  {Watanabe}},\ }\href {\doibase 10.1103/PhysRevLett.96.255503} {\bibfield
  {journal} {\bibinfo  {journal} {Phys. Rev. Lett.}\ }\textbf {\bibinfo
  {volume} {96}},\ \bibinfo {pages} {255503} (\bibinfo {year}
  {2006})}\BibitemShut {NoStop}%
\bibitem [{\citenamefont {Wang}\ \emph {et~al.}(2006)\citenamefont {Wang},
  \citenamefont {Wang},\ and\ \citenamefont {Zeng}}]{wang2006nonequilibrium}%
  \BibitemOpen
  \bibfield  {author} {\bibinfo {author} {\bibfnamefont {J.-S.}\ \bibnamefont
  {Wang}}, \bibinfo {author} {\bibfnamefont {J.}~\bibnamefont {Wang}}, \ and\
  \bibinfo {author} {\bibfnamefont {N.}~\bibnamefont {Zeng}},\ }\href {\doibase
  10.1103/PhysRevB.74.033408} {\bibfield  {journal} {\bibinfo  {journal} {Phys.
  Rev. B}\ }\textbf {\bibinfo {volume} {74}},\ \bibinfo {pages} {033408}
  (\bibinfo {year} {2006})}\BibitemShut {NoStop}%
\bibitem [{\citenamefont {Wang}\ \emph {et~al.}(2007)\citenamefont {Wang},
  \citenamefont {Zeng}, \citenamefont {Wang},\ and\ \citenamefont
  {Gan}}]{wang2007nonequilibrium}%
  \BibitemOpen
  \bibfield  {author} {\bibinfo {author} {\bibfnamefont {J.-S.}\ \bibnamefont
  {Wang}}, \bibinfo {author} {\bibfnamefont {N.}~\bibnamefont {Zeng}}, \bibinfo
  {author} {\bibfnamefont {J.}~\bibnamefont {Wang}}, \ and\ \bibinfo {author}
  {\bibfnamefont {C.~K.}\ \bibnamefont {Gan}},\ }\href {\doibase
  10.1103/PhysRevE.75.061128} {\bibfield  {journal} {\bibinfo  {journal} {Phys.
  Rev. E}\ }\textbf {\bibinfo {volume} {75}},\ \bibinfo {pages} {061128}
  (\bibinfo {year} {2007})}\BibitemShut {NoStop}%
\bibitem [{\citenamefont {Wang}\ \emph {et~al.}(2008)\citenamefont {Wang},
  \citenamefont {Wang},\ and\ \citenamefont {L{{\"u}}}}]{wang2008quantum}%
  \BibitemOpen
  \bibfield  {author} {\bibinfo {author} {\bibfnamefont {J.-S.}\ \bibnamefont
  {Wang}}, \bibinfo {author} {\bibfnamefont {J.}~\bibnamefont {Wang}}, \ and\
  \bibinfo {author} {\bibfnamefont {J.-T.}\ \bibnamefont {L{{\"u}}}},\ }\href
  {\doibase 10.1140/epjb/e2008-00195-8} {\bibfield  {journal} {\bibinfo
  {journal} {Eur. Phys. J. B}\ }\textbf {\bibinfo {volume} {62}},\ \bibinfo
  {pages} {381} (\bibinfo {year} {2008})}\BibitemShut {NoStop}%
\bibitem [{\citenamefont {Ruokola}\ \emph {et~al.}(2009)\citenamefont
  {Ruokola}, \citenamefont {Ojanen},\ and\ \citenamefont
  {Jauho}}]{ruokola2009thermal}%
  \BibitemOpen
  \bibfield  {author} {\bibinfo {author} {\bibfnamefont {T.}~\bibnamefont
  {Ruokola}}, \bibinfo {author} {\bibfnamefont {T.}~\bibnamefont {Ojanen}}, \
  and\ \bibinfo {author} {\bibfnamefont {A.-P.}\ \bibnamefont {Jauho}},\ }\href
  {\doibase 10.1103/PhysRevB.79.144306} {\bibfield  {journal} {\bibinfo
  {journal} {Phys. Rev. B}\ }\textbf {\bibinfo {volume} {79}},\ \bibinfo
  {pages} {144306} (\bibinfo {year} {2009})}\BibitemShut {NoStop}%
\bibitem [{\citenamefont {Li}\ \emph {et~al.}(2012)\citenamefont {Li},
  \citenamefont {Ren}, \citenamefont {Wang}, \citenamefont {Zhang},
  \citenamefont {H{\"a}nggi},\ and\ \citenamefont {Li}}]{li2012colloquium}%
  \BibitemOpen
  \bibfield  {author} {\bibinfo {author} {\bibfnamefont {N.}~\bibnamefont
  {Li}}, \bibinfo {author} {\bibfnamefont {J.}~\bibnamefont {Ren}}, \bibinfo
  {author} {\bibfnamefont {L.}~\bibnamefont {Wang}}, \bibinfo {author}
  {\bibfnamefont {G.}~\bibnamefont {Zhang}}, \bibinfo {author} {\bibfnamefont
  {P.}~\bibnamefont {H{\"a}nggi}}, \ and\ \bibinfo {author} {\bibfnamefont
  {B.}~\bibnamefont {Li}},\ }\href
  {https://journals.aps.org/rmp/abstract/10.1103/RevModPhys.84.1419} {\bibfield
   {journal} {\bibinfo  {journal} {Rev. Mod. Phys.}\ }\textbf {\bibinfo
  {volume} {84}},\ \bibinfo {pages} {1045} (\bibinfo {year}
  {2012})}\BibitemShut {NoStop}%
\bibitem [{\citenamefont {Taylor}\ and\ \citenamefont
  {Segal}(2015)}]{taylor2015quantum}%
  \BibitemOpen
  \bibfield  {author} {\bibinfo {author} {\bibfnamefont {E.}~\bibnamefont
  {Taylor}}\ and\ \bibinfo {author} {\bibfnamefont {D.}~\bibnamefont {Segal}},\
  }\href {\doibase 10.1103/PhysRevLett.114.220401} {\bibfield  {journal}
  {\bibinfo  {journal} {Phys. Rev. Lett.}\ }\textbf {\bibinfo {volume} {114}},\
  \bibinfo {pages} {220401} (\bibinfo {year} {2015})}\BibitemShut {NoStop}%
\bibitem [{\citenamefont {Wang}\ and\ \citenamefont
  {Taylor}(2016)}]{wang2016landauer}%
  \BibitemOpen
  \bibfield  {author} {\bibinfo {author} {\bibfnamefont {C.-H.}\ \bibnamefont
  {Wang}}\ and\ \bibinfo {author} {\bibfnamefont {J.~M.}\ \bibnamefont
  {Taylor}},\ }\href {\doibase 10.1103/PhysRevB.94.155437} {\bibfield
  {journal} {\bibinfo  {journal} {Phys. Rev. B}\ }\textbf {\bibinfo {volume}
  {94}},\ \bibinfo {pages} {155437} (\bibinfo {year} {2016})}\BibitemShut
  {NoStop}%
\bibitem [{\citenamefont {Wang}\ \emph {et~al.}(2004)\citenamefont {Wang},
  \citenamefont {Wang}, \citenamefont {Wang},\ and\ \citenamefont
  {Xing}}]{wang2004spin}%
  \BibitemOpen
  \bibfield  {author} {\bibinfo {author} {\bibfnamefont {B.}~\bibnamefont
  {Wang}}, \bibinfo {author} {\bibfnamefont {J.}~\bibnamefont {Wang}}, \bibinfo
  {author} {\bibfnamefont {J.}~\bibnamefont {Wang}}, \ and\ \bibinfo {author}
  {\bibfnamefont {D.~Y.}\ \bibnamefont {Xing}},\ }\href {\doibase
  10.1103/PhysRevB.69.174403} {\bibfield  {journal} {\bibinfo  {journal} {Phys.
  Rev. B}\ }\textbf {\bibinfo {volume} {69}},\ \bibinfo {pages} {174403}
  (\bibinfo {year} {2004})}\BibitemShut {NoStop}%
\bibitem [{\citenamefont {Kuhnke}\ \emph {et~al.}(2017)\citenamefont {Kuhnke},
  \citenamefont {Gro{\ss}e}, \citenamefont {Merino},\ and\ \citenamefont
  {Kern}}]{kuhnke2017atomic}%
  \BibitemOpen
  \bibfield  {author} {\bibinfo {author} {\bibfnamefont {K.}~\bibnamefont
  {Kuhnke}}, \bibinfo {author} {\bibfnamefont {C.}~\bibnamefont {Gro{\ss}e}},
  \bibinfo {author} {\bibfnamefont {P.}~\bibnamefont {Merino}}, \ and\ \bibinfo
  {author} {\bibfnamefont {K.}~\bibnamefont {Kern}},\ }\href {\doibase
  10.1021/acs.chemrev.6b00645} {\bibfield  {journal} {\bibinfo  {journal}
  {Chem. Rev.}\ }\textbf {\bibinfo {volume} {117}},\ \bibinfo {pages} {5174}
  (\bibinfo {year} {2017})}\BibitemShut {NoStop}%
\bibitem [{\citenamefont {Galperin}(2017)}]{galeprin2017photonics}%
  \BibitemOpen
  \bibfield  {author} {\bibinfo {author} {\bibfnamefont {M.}~\bibnamefont
  {Galperin}},\ }\href {\doibase 10.1039/C7CS00067G} {\bibfield  {journal}
  {\bibinfo  {journal} {Chem. Soc. Rev.}\ }\textbf {\bibinfo {volume} {46}},\
  \bibinfo {pages} {4000} (\bibinfo {year} {2017})}\BibitemShut {NoStop}%
\bibitem [{\citenamefont {Schneider}\ \emph {et~al.}(2010)\citenamefont
  {Schneider}, \citenamefont {Schull},\ and\ \citenamefont
  {Berndt}}]{schneider2010optical}%
  \BibitemOpen
  \bibfield  {author} {\bibinfo {author} {\bibfnamefont {N.~L.}\ \bibnamefont
  {Schneider}}, \bibinfo {author} {\bibfnamefont {G.}~\bibnamefont {Schull}}, \
  and\ \bibinfo {author} {\bibfnamefont {R.}~\bibnamefont {Berndt}},\ }\href
  {\doibase 10.1103/PhysRevLett.105.026601} {\bibfield  {journal} {\bibinfo
  {journal} {Phys. Rev. Lett.}\ }\textbf {\bibinfo {volume} {105}},\ \bibinfo
  {pages} {026601} (\bibinfo {year} {2010})}\BibitemShut {NoStop}%
\bibitem [{\citenamefont {Schneider}\ \emph {et~al.}(2012)\citenamefont
  {Schneider}, \citenamefont {L{\"u}}, \citenamefont {Brandbyge},\ and\
  \citenamefont {Berndt}}]{schneider2012light}%
  \BibitemOpen
  \bibfield  {author} {\bibinfo {author} {\bibfnamefont {N.~L.}\ \bibnamefont
  {Schneider}}, \bibinfo {author} {\bibfnamefont {J.-T.}\ \bibnamefont
  {L{\"u}}}, \bibinfo {author} {\bibfnamefont {M.}~\bibnamefont {Brandbyge}}, \
  and\ \bibinfo {author} {\bibfnamefont {R.}~\bibnamefont {Berndt}},\ }\href
  {\doibase 10.1103/PhysRevLett.109.186601} {\bibfield  {journal} {\bibinfo
  {journal} {Phys. Rev. Lett.}\ }\textbf {\bibinfo {volume} {109}},\ \bibinfo
  {pages} {186601} (\bibinfo {year} {2012})}\BibitemShut {NoStop}%
\bibitem [{\citenamefont {Huang}\ \emph {et~al.}(2007)\citenamefont {Huang},
  \citenamefont {Chen}, \citenamefont {D'agosta}, \citenamefont {Bennett},
  \citenamefont {Di~Ventra},\ and\ \citenamefont {Tao}}]{huang2007local}%
  \BibitemOpen
  \bibfield  {author} {\bibinfo {author} {\bibfnamefont {Z.}~\bibnamefont
  {Huang}}, \bibinfo {author} {\bibfnamefont {F.}~\bibnamefont {Chen}},
  \bibinfo {author} {\bibfnamefont {R.}~\bibnamefont {D'agosta}}, \bibinfo
  {author} {\bibfnamefont {P.~A.}\ \bibnamefont {Bennett}}, \bibinfo {author}
  {\bibfnamefont {M.}~\bibnamefont {Di~Ventra}}, \ and\ \bibinfo {author}
  {\bibfnamefont {N.}~\bibnamefont {Tao}},\ }\href
  {https://www.nature.com/articles/nnano.2007.345} {\bibfield  {journal}
  {\bibinfo  {journal} {Nat. Nanotechnol.}\ }\textbf {\bibinfo {volume} {2}},\
  \bibinfo {pages} {698} (\bibinfo {year} {2007})}\BibitemShut {NoStop}%
\bibitem [{\citenamefont {Ioffe}\ \emph {et~al.}(2008)\citenamefont {Ioffe},
  \citenamefont {Shamai}, \citenamefont {Ophir}, \citenamefont {Noy},
  \citenamefont {Yutsis}, \citenamefont {Kfir}, \citenamefont {Cheshnovsky},\
  and\ \citenamefont {Selzer}}]{ioffe2008detection}%
  \BibitemOpen
  \bibfield  {author} {\bibinfo {author} {\bibfnamefont {Z.}~\bibnamefont
  {Ioffe}}, \bibinfo {author} {\bibfnamefont {T.}~\bibnamefont {Shamai}},
  \bibinfo {author} {\bibfnamefont {A.}~\bibnamefont {Ophir}}, \bibinfo
  {author} {\bibfnamefont {G.}~\bibnamefont {Noy}}, \bibinfo {author}
  {\bibfnamefont {I.}~\bibnamefont {Yutsis}}, \bibinfo {author} {\bibfnamefont
  {K.}~\bibnamefont {Kfir}}, \bibinfo {author} {\bibfnamefont {O.}~\bibnamefont
  {Cheshnovsky}}, \ and\ \bibinfo {author} {\bibfnamefont {Y.}~\bibnamefont
  {Selzer}},\ }\href {https://www.nature.com/articles/nnano.2008.304}
  {\bibfield  {journal} {\bibinfo  {journal} {Nat. Nanotechnol.}\ }\textbf
  {\bibinfo {volume} {3}},\ \bibinfo {pages} {727} (\bibinfo {year}
  {2008})}\BibitemShut {NoStop}%
\bibitem [{\citenamefont {L{\"u}}\ \emph {et~al.}(2015)\citenamefont {L{\"u}},
  \citenamefont {Christensen}, \citenamefont {Wang}, \citenamefont
  {Hedeg{\aa}rd},\ and\ \citenamefont {Brandbyge}}]{lu_current-induced_2015}%
  \BibitemOpen
  \bibfield  {author} {\bibinfo {author} {\bibfnamefont {J.-T.}\ \bibnamefont
  {L{\"u}}}, \bibinfo {author} {\bibfnamefont {R.~B.}\ \bibnamefont
  {Christensen}}, \bibinfo {author} {\bibfnamefont {J.-S.}\ \bibnamefont
  {Wang}}, \bibinfo {author} {\bibfnamefont {P.}~\bibnamefont {Hedeg{\aa}rd}},
  \ and\ \bibinfo {author} {\bibfnamefont {M.}~\bibnamefont {Brandbyge}},\
  }\href {\doibase 10.1103/PhysRevLett.114.096801} {\bibfield  {journal}
  {\bibinfo  {journal} {Phys. Rev. Lett.}\ }\textbf {\bibinfo {volume} {114}},\
  \bibinfo {pages} {096801} (\bibinfo {year} {2015})}\BibitemShut {NoStop}%
\bibitem [{\citenamefont {H{\"a}rtle}\ and\ \citenamefont
  {Thoss}(2011{\natexlab{a}})}]{hartle2011resonant}%
  \BibitemOpen
  \bibfield  {author} {\bibinfo {author} {\bibfnamefont {R.}~\bibnamefont
  {H{\"a}rtle}}\ and\ \bibinfo {author} {\bibfnamefont {M.}~\bibnamefont
  {Thoss}},\ }\href {\doibase 10.1103/PhysRevB.83.115414} {\bibfield  {journal}
  {\bibinfo  {journal} {Phys. Rev. B}\ }\textbf {\bibinfo {volume} {83}},\
  \bibinfo {pages} {115414} (\bibinfo {year} {2011}{\natexlab{a}})}\BibitemShut
  {NoStop}%
\bibitem [{\citenamefont {H{\"a}rtle}\ and\ \citenamefont
  {Thoss}(2011{\natexlab{b}})}]{hartle2011vibrational}%
  \BibitemOpen
  \bibfield  {author} {\bibinfo {author} {\bibfnamefont {R.}~\bibnamefont
  {H{\"a}rtle}}\ and\ \bibinfo {author} {\bibfnamefont {M.}~\bibnamefont
  {Thoss}},\ }\href {\doibase 10.1103/PhysRevB.83.125419} {\bibfield  {journal}
  {\bibinfo  {journal} {Phys. Rev. B}\ }\textbf {\bibinfo {volume} {83}},\
  \bibinfo {pages} {125419} (\bibinfo {year} {2011}{\natexlab{b}})}\BibitemShut
  {NoStop}%
\bibitem [{\citenamefont {H{\"a}rtle}\ \emph {et~al.}(2018)\citenamefont
  {H{\"a}rtle}, \citenamefont {Schinabeck}, \citenamefont {Kulkarni},
  \citenamefont {Gelbwaser-Klimovsky}, \citenamefont {Thoss},\ and\
  \citenamefont {Peskin}}]{hartle2018cooling}%
  \BibitemOpen
  \bibfield  {author} {\bibinfo {author} {\bibfnamefont {R.}~\bibnamefont
  {H{\"a}rtle}}, \bibinfo {author} {\bibfnamefont {C.}~\bibnamefont
  {Schinabeck}}, \bibinfo {author} {\bibfnamefont {M.}~\bibnamefont
  {Kulkarni}}, \bibinfo {author} {\bibfnamefont {D.}~\bibnamefont
  {Gelbwaser-Klimovsky}}, \bibinfo {author} {\bibfnamefont {M.}~\bibnamefont
  {Thoss}}, \ and\ \bibinfo {author} {\bibfnamefont {U.}~\bibnamefont
  {Peskin}},\ }\href {\doibase 10.1103/PhysRevB.98.081404} {\bibfield
  {journal} {\bibinfo  {journal} {Phys. Rev. B}\ }\textbf {\bibinfo {volume}
  {98}},\ \bibinfo {pages} {081404} (\bibinfo {year} {2018})}\BibitemShut
  {NoStop}%
\bibitem [{\citenamefont {Galperin}\ \emph {et~al.}(2009)\citenamefont
  {Galperin}, \citenamefont {Saito}, \citenamefont {Balatsky},\ and\
  \citenamefont {Nitzan}}]{galperin2009cooling}%
  \BibitemOpen
  \bibfield  {author} {\bibinfo {author} {\bibfnamefont {M.}~\bibnamefont
  {Galperin}}, \bibinfo {author} {\bibfnamefont {K.}~\bibnamefont {Saito}},
  \bibinfo {author} {\bibfnamefont {A.~V.}\ \bibnamefont {Balatsky}}, \ and\
  \bibinfo {author} {\bibfnamefont {A.}~\bibnamefont {Nitzan}},\ }\href
  {\doibase 10.1103/PhysRevB.80.115427} {\bibfield  {journal} {\bibinfo
  {journal} {Phys. Rev. B}\ }\textbf {\bibinfo {volume} {80}},\ \bibinfo
  {pages} {115427} (\bibinfo {year} {2009})}\BibitemShut {NoStop}%
\bibitem [{\citenamefont {Simine}\ and\ \citenamefont
  {Segal}(2012)}]{simine2012vibrational}%
  \BibitemOpen
  \bibfield  {author} {\bibinfo {author} {\bibfnamefont {L.}~\bibnamefont
  {Simine}}\ and\ \bibinfo {author} {\bibfnamefont {D.}~\bibnamefont {Segal}},\
  }\href
  {https://pubs.rsc.org/en/content/articlelanding/2012/CP/c2cp40851a#!divAbstract}
  {\bibfield  {journal} {\bibinfo  {journal} {Phys. Chem. Chem. Phys.}\
  }\textbf {\bibinfo {volume} {14}},\ \bibinfo {pages} {13820} (\bibinfo {year}
  {2012})}\BibitemShut {NoStop}%
\bibitem [{\citenamefont {Lykkebo}\ \emph {et~al.}(2016)\citenamefont
  {Lykkebo}, \citenamefont {Romano}, \citenamefont {Gagliardi}, \citenamefont
  {Pecchia},\ and\ \citenamefont {Solomon}}]{lykkebo2016single}%
  \BibitemOpen
  \bibfield  {author} {\bibinfo {author} {\bibfnamefont {J.}~\bibnamefont
  {Lykkebo}}, \bibinfo {author} {\bibfnamefont {G.}~\bibnamefont {Romano}},
  \bibinfo {author} {\bibfnamefont {A.}~\bibnamefont {Gagliardi}}, \bibinfo
  {author} {\bibfnamefont {A.}~\bibnamefont {Pecchia}}, \ and\ \bibinfo
  {author} {\bibfnamefont {G.~C.}\ \bibnamefont {Solomon}},\ }\href {\doibase
  10.1063/1.4943578} {\bibfield  {journal} {\bibinfo  {journal} {J. Chem.
  Phys.}\ }\textbf {\bibinfo {volume} {144}},\ \bibinfo {pages} {114310}
  (\bibinfo {year} {2016})}\BibitemShut {NoStop}%
\bibitem [{\citenamefont {Zhu}\ \emph {et~al.}(2019)\citenamefont {Zhu},
  \citenamefont {Fiorino}, \citenamefont {Thompson}, \citenamefont
  {Mittapally}, \citenamefont {Meyhofer},\ and\ \citenamefont
  {Reddy}}]{zhu2019near}%
  \BibitemOpen
  \bibfield  {author} {\bibinfo {author} {\bibfnamefont {L.}~\bibnamefont
  {Zhu}}, \bibinfo {author} {\bibfnamefont {A.}~\bibnamefont {Fiorino}},
  \bibinfo {author} {\bibfnamefont {D.}~\bibnamefont {Thompson}}, \bibinfo
  {author} {\bibfnamefont {R.}~\bibnamefont {Mittapally}}, \bibinfo {author}
  {\bibfnamefont {E.}~\bibnamefont {Meyhofer}}, \ and\ \bibinfo {author}
  {\bibfnamefont {P.}~\bibnamefont {Reddy}},\ }\href
  {https://www.nature.com/articles/s41586-019-0918-8} {\bibfield  {journal}
  {\bibinfo  {journal} {Nature}\ }\textbf {\bibinfo {volume} {566}},\ \bibinfo
  {pages} {239} (\bibinfo {year} {2019})}\BibitemShut {NoStop}%
\bibitem [{\citenamefont {Head‐Gordon}\ and\ \citenamefont
  {Tully}(1995)}]{headgordon_molecular_1995}%
  \BibitemOpen
  \bibfield  {author} {\bibinfo {author} {\bibfnamefont {M.}~\bibnamefont
  {Head‐Gordon}}\ and\ \bibinfo {author} {\bibfnamefont {J.~C.}\ \bibnamefont
  {Tully}},\ }\href {\doibase 10.1063/1.469915} {\bibfield  {journal} {\bibinfo
   {journal} {J. Chem. Phys.}\ }\textbf {\bibinfo {volume} {103}},\ \bibinfo
  {pages} {10137} (\bibinfo {year} {1995})}\BibitemShut {NoStop}%
\bibitem [{\citenamefont {Dou}\ and\ \citenamefont
  {Subotnik}(2018)}]{dou2018perspective}%
  \BibitemOpen
  \bibfield  {author} {\bibinfo {author} {\bibfnamefont {W.}~\bibnamefont
  {Dou}}\ and\ \bibinfo {author} {\bibfnamefont {J.~E.}\ \bibnamefont
  {Subotnik}},\ }\href {\doibase 10.1063/1.5035412} {\bibfield  {journal}
  {\bibinfo  {journal} {J. Chem. Phys.}\ }\textbf {\bibinfo {volume} {148}},\
  \bibinfo {pages} {230901} (\bibinfo {year} {2018})}\BibitemShut {NoStop}%
\bibitem [{\citenamefont {L{\"u}}\ \emph {et~al.}(2012)\citenamefont {L{\"u}},
  \citenamefont {Brandbyge}, \citenamefont {Hedeg{\aa}rd}, \citenamefont
  {Todorov},\ and\ \citenamefont {Dundas}}]{lu_current-induced_2012}%
  \BibitemOpen
  \bibfield  {author} {\bibinfo {author} {\bibfnamefont {J.-T.}\ \bibnamefont
  {L{\"u}}}, \bibinfo {author} {\bibfnamefont {M.}~\bibnamefont {Brandbyge}},
  \bibinfo {author} {\bibfnamefont {P.}~\bibnamefont {Hedeg{\aa}rd}}, \bibinfo
  {author} {\bibfnamefont {T.~N.}\ \bibnamefont {Todorov}}, \ and\ \bibinfo
  {author} {\bibfnamefont {D.}~\bibnamefont {Dundas}},\ }\href {\doibase
  10.1103/PhysRevB.85.245444} {\bibfield  {journal} {\bibinfo  {journal} {Phys.
  Rev. B}\ }\textbf {\bibinfo {volume} {85}},\ \bibinfo {pages} {245444}
  (\bibinfo {year} {2012})}\BibitemShut {NoStop}%
\bibitem [{\citenamefont {L{{\"u}}}\ \emph {et~al.}(2016)\citenamefont
  {L{{\"u}}}, \citenamefont {Wang}, \citenamefont {Hedeg{\aa}rd},\ and\
  \citenamefont {Brandbyge}}]{lu2016electron}%
  \BibitemOpen
  \bibfield  {author} {\bibinfo {author} {\bibfnamefont {J.-T.}\ \bibnamefont
  {L{{\"u}}}}, \bibinfo {author} {\bibfnamefont {J.-S.}\ \bibnamefont {Wang}},
  \bibinfo {author} {\bibfnamefont {P.}~\bibnamefont {Hedeg{\aa}rd}}, \ and\
  \bibinfo {author} {\bibfnamefont {M.}~\bibnamefont {Brandbyge}},\ }\href
  {\doibase 10.1103/PhysRevB.93.205404} {\bibfield  {journal} {\bibinfo
  {journal} {Phys. Rev. B}\ }\textbf {\bibinfo {volume} {93}},\ \bibinfo
  {pages} {205404} (\bibinfo {year} {2016})}\BibitemShut {NoStop}%
\bibitem [{\citenamefont {L{{\"u}}}\ \emph {et~al.}(2011)\citenamefont
  {L{{\"u}}}, \citenamefont {Hedeg{\aa}rd},\ and\ \citenamefont
  {Brandbyge}}]{lu2011laserlike}%
  \BibitemOpen
  \bibfield  {author} {\bibinfo {author} {\bibfnamefont {J.-T.}\ \bibnamefont
  {L{{\"u}}}}, \bibinfo {author} {\bibfnamefont {P.}~\bibnamefont
  {Hedeg{\aa}rd}}, \ and\ \bibinfo {author} {\bibfnamefont {M.}~\bibnamefont
  {Brandbyge}},\ }\href {\doibase 10.1103/PhysRevLett.107.046801} {\bibfield
  {journal} {\bibinfo  {journal} {Phys. Rev. Lett.}\ }\textbf {\bibinfo
  {volume} {107}},\ \bibinfo {pages} {046801} (\bibinfo {year}
  {2011})}\BibitemShut {NoStop}%
\bibitem [{\citenamefont {Nitzan}\ and\ \citenamefont
  {Galperin}(2018)}]{nitzan2018kinetic}%
  \BibitemOpen
  \bibfield  {author} {\bibinfo {author} {\bibfnamefont {A.}~\bibnamefont
  {Nitzan}}\ and\ \bibinfo {author} {\bibfnamefont {M.}~\bibnamefont
  {Galperin}},\ }\href {\doibase 10.1021/acs.jpclett.8b01886} {\bibfield
  {journal} {\bibinfo  {journal} {J. Phys. Chem. Lett.}\ }\textbf {\bibinfo
  {volume} {9}},\ \bibinfo {pages} {4886} (\bibinfo {year} {2018})}\BibitemShut
  {NoStop}%
\bibitem [{\citenamefont {Wang}\ \emph {et~al.}()\citenamefont {Wang},
  \citenamefont {Nian},\ and\ \citenamefont {L\"u}}]{Wang-preprint}%
  \BibitemOpen
  \bibfield  {author} {\bibinfo {author} {\bibfnamefont {T.}~\bibnamefont
  {Wang}}, \bibinfo {author} {\bibfnamefont {L.-L.}\ \bibnamefont {Nian}}, \
  and\ \bibinfo {author} {\bibfnamefont {J.-T.}\ \bibnamefont {L\"u}},\ }\href
  {https://arxiv.org/abs/2003.09614} {\ }\Eprint
  {http://arxiv.org/abs/2003.09614} {arXiv:2003.09614 [cond-mat.mes-hall]}
  \BibitemShut {NoStop}%
\bibitem [{\citenamefont {Zhang}\ \emph {et~al.}(2013)\citenamefont {Zhang},
  \citenamefont {L{{\"u}}}, \citenamefont {Wang},\ and\ \citenamefont
  {Li}}]{zhang2013thermal}%
  \BibitemOpen
  \bibfield  {author} {\bibinfo {author} {\bibfnamefont {L.}~\bibnamefont
  {Zhang}}, \bibinfo {author} {\bibfnamefont {J.-T.}\ \bibnamefont {L{{\"u}}}},
  \bibinfo {author} {\bibfnamefont {J.-S.}\ \bibnamefont {Wang}}, \ and\
  \bibinfo {author} {\bibfnamefont {B.}~\bibnamefont {Li}},\ }\href
  {https://iopscience.iop.org/article/10.1088/0953-8984/25/44/445801/meta}
  {\bibfield  {journal} {\bibinfo  {journal} {J. Phys.: Condens. Matt.}\
  }\textbf {\bibinfo {volume} {25}},\ \bibinfo {pages} {445801} (\bibinfo
  {year} {2013})}\BibitemShut {NoStop}%
\bibitem [{\citenamefont {Ren}\ and\ \citenamefont {Zhu}(2013)}]{ren2013heat}%
  \BibitemOpen
  \bibfield  {author} {\bibinfo {author} {\bibfnamefont {J.}~\bibnamefont
  {Ren}}\ and\ \bibinfo {author} {\bibfnamefont {J.-X.}\ \bibnamefont {Zhu}},\
  }\href {\doibase 10.1103/PhysRevB.87.241412} {\bibfield  {journal} {\bibinfo
  {journal} {Phys. Rev. B}\ }\textbf {\bibinfo {volume} {87}},\ \bibinfo
  {pages} {241412} (\bibinfo {year} {2013})}\BibitemShut {NoStop}%
\bibitem [{\citenamefont {Entin-Wohlman}\ \emph {et~al.}(2010)\citenamefont
  {Entin-Wohlman}, \citenamefont {Imry},\ and\ \citenamefont
  {Aharony}}]{entinwuhlman2010three}%
  \BibitemOpen
  \bibfield  {author} {\bibinfo {author} {\bibfnamefont {O.}~\bibnamefont
  {Entin-Wohlman}}, \bibinfo {author} {\bibfnamefont {Y.}~\bibnamefont {Imry}},
  \ and\ \bibinfo {author} {\bibfnamefont {A.}~\bibnamefont {Aharony}},\ }\href
  {\doibase 10.1103/PhysRevB.82.115314} {\bibfield  {journal} {\bibinfo
  {journal} {Phys. Rev. B}\ }\textbf {\bibinfo {volume} {82}},\ \bibinfo
  {pages} {115314} (\bibinfo {year} {2010})}\BibitemShut {NoStop}%
\bibitem [{\citenamefont {S\'anchez}\ and\ \citenamefont
  {B{\"u}ttiker}(2011)}]{sanchez2011optimal}%
  \BibitemOpen
  \bibfield  {author} {\bibinfo {author} {\bibfnamefont {R.}~\bibnamefont
  {S\'anchez}}\ and\ \bibinfo {author} {\bibfnamefont {M.}~\bibnamefont
  {B{\"u}ttiker}},\ }\href {\doibase 10.1103/PhysRevB.83.085428} {\bibfield
  {journal} {\bibinfo  {journal} {Phys. Rev. B}\ }\textbf {\bibinfo {volume}
  {83}},\ \bibinfo {pages} {085428} (\bibinfo {year} {2011})}\BibitemShut
  {NoStop}%
\end{thebibliography}%
\bibliographystyle{apsrev4-1}
\end{document}